\documentclass[floats,twocolumn,pra]{revtex4}
\usepackage{eurosym}
\usepackage{graphics}
\usepackage[thmmarks]{ntheorem}
\usepackage{graphicx}
\usepackage{algorithm}
\usepackage{algpseudocode}
\usepackage{amsmath}
\usepackage{amsfonts}
\usepackage{amssymb}
\usepackage{float}
\usepackage{longtable}
\usepackage{flushend}
\usepackage{epsfig}
\usepackage{latexsym}
\usepackage{theorem}
\usepackage{bbm}
\usepackage{bm}
\usepackage{psfrag}
\usepackage{eurosym}
\usepackage[utf8]{inputenc}
\usepackage[T1]{fontenc}
\usepackage{physics}%
\usepackage{color}
\usepackage{multirow}
\usepackage{float}
\usepackage{soul}
\usepackage{ulem}
\usepackage{cancel}
\usepackage{hyperref}
\hypersetup{
    colorlinks=true,
    linkcolor=blue,
    filecolor=magenta,
    urlcolor=cyan,
}

\DeclareMathOperator*{\argmax}{arg\,max}
\begin{document}
\title{Composably secure data processing for  \\Gaussian-modulated continuous variable quantum key distribution}
\author{Alexander G. Mountogiannakis$^{1}$}
\author{Panagiotis Papanastasiou$^{1}$}
\author{Boris Braverman$^{2}$}
\author{Stefano Pirandola$^{1}$}
\affiliation{$^1$Department of Computer Science, University of York, York YO10 5GH, UK
\\
$^2$Xanadu, 372 Richmond St W, Toronto, M5V 2L7, Canada}

\begin{abstract}
Continuous-variable (CV) quantum key distribution (QKD) employs the quadratures of a bosonic mode to establish a secret key between two remote parties, and this is usually achieved via a Gaussian modulation of coherent states. The
resulting secret key rate depends not only on the loss and noise in the communication channel,
but also on a series of data processing steps that are
needed for transforming shared correlations into a final string of secret
bits. Here we consider a Gaussian-modulated coherent-state protocol with homodyne detection in the general setting of composable finite-size security. After simulating the process of quantum communication, the output classical data is post-processed via procedures of parameter estimation,
error correction, and privacy amplification. 
In particular, we analyze the high signal-to-noise regime which requires the use of high-rate (non-binary) low-density parity check codes.
We implement all these steps in a Python-based library that allows one to investigate and optimize the protocol parameters to be used in practical experimental implementations of short-range CV-QKD.
\end{abstract}

\maketitle

\section{Introduction}

In quantum key distribution (QKD), two authenticated
parties (Alice and Bob) aim at establishing a secret key over a potentially insecure quantum channel~\cite{revQKD}.
The security of QKD is derived from the laws of quantum mechanics, namely the uncertainty
principle or the no-cloning theorem~\cite{clone1,clone2}. In 1984, Bennett and Brassard introduced
the first QKD protocol~\cite{BB84}, in which information is encoded on a property of a
discrete-variable (DV) system, such as the polarization of a
photon. This class of protocols started the area of DV-QKD and their security has been extensively
studied~\cite{revQKD, Darius}.

Later on, at the beginning of the 2000s, a more modern family of protocols emerged, in which information is
encoded in the position and momentum quadratures of a bosonic mode~\cite{GG02,noswitch}. These
continuous-variable (CV) QKD protocols are particularly practical and
cost-effective, being already compatible with the current telecommunication technology~\cite{revQKD,Gaussian_rev}.
CV-QKD protocols are very suitable to reach high rates of communication
(e.g. comparable with the ultimate channel limits~\cite{PLOB}) over distances that are compatible with metropolitan areas~\cite{CVMDI}.
Recently, CV-QKD protocols have also been shown to reach very long distances, comparable to those of DV protocols~\cite{LeoEXP,LeoCodes}.

In a basic CV-QKD protocol, Alice encodes a classical variable in the coherent
states of a bosonic mode via Gaussian modulation. These states are then
transmitted to Bob via the noisy communication channel. At the output, Bob
measures the received states via homodyne~\cite{GG02,LeoEXP} (or heterodyne~\cite{noswitch}) detection in order to
retrieve Alice's encoded information. In the worst-case scenario,
all the loss and noise present in the channel is ascribed to a malicious party
(Eve) who tries to intercept the states and obtain information about the key.
By carefully estimating Eve's perturbation, Alice and Bob can apply procedures
of error correction (EC) to remove noise from their data, and then privacy
amplification (PA), which makes their error-corrected data completely secret.

In this work, we consider the basic CV-QKD protocol based on Gaussian modulation
of coherent states and homodyne detection. 
Starting from a simulation of the quantum communication in typical noisy conditions, we process the generated data
into a finite-size secret key which is composably secure against collective Gaussian attacks.
Besides developing the technical procedure step-by-step, we implement it in an open-access Python library associated with this paper~\cite{library}. In particular, our procedure for data processing is based on the composable secret key rate developed in Ref.~\cite[Sec. III]{freeSPACE}.

In order to address a short-range regime with relatively high values of the signal to noise ratio (SNR), the step of EC
is based on high-rate (non-binary) low-density parity check codes (LDPC)~\cite{Mackay1, Mackay2,Duan,Duan2,Pacher_LDPC}, whose
decoding is performed by means of a suitable iterative sum-product algorithm~\cite{Mackay2,sumALGO}
(even though the min-sum algorithm~\cite{Mario} can be used as an alternative). 
After EC, the procedure of PA is based on universal hash functions~\cite{hashAPP}.
Because of the length of the generated strings can be substantial, techniques with
low complexity are preferred, so that we use the Toeplitz-based hash
function, which is simple, parallelizable and can moreover be accelerated by using the
Fast Fourier Transform (FFT)~\cite{FFT}.

The paper is structured as follows. In Sec.~\ref{sec:coherentanalysis} we start with the
description of the coherent-state protocol including a realistic model for the detector setup.
After an initial analysis of its asymptotic security, we discuss the steps of parameter
estimation (PE), EC and PA, ending with the composable secret key rate of the protocol. In Sec.~\ref{sec:processing},
we go into further details of the protocol simulation and data processing, also presenting the pseudocode
for the entire process. In Sec.~\ref{sec:simulations} we provide numerical results that are obtained by
our Python library in simulated experiments. Finally, Sec.~\ref{sec:conclusions} is for conclusions.

\begin{figure}[t]
\vspace{-1.3cm}
\par
\begin{center}
\includegraphics[width=0.5\textwidth] {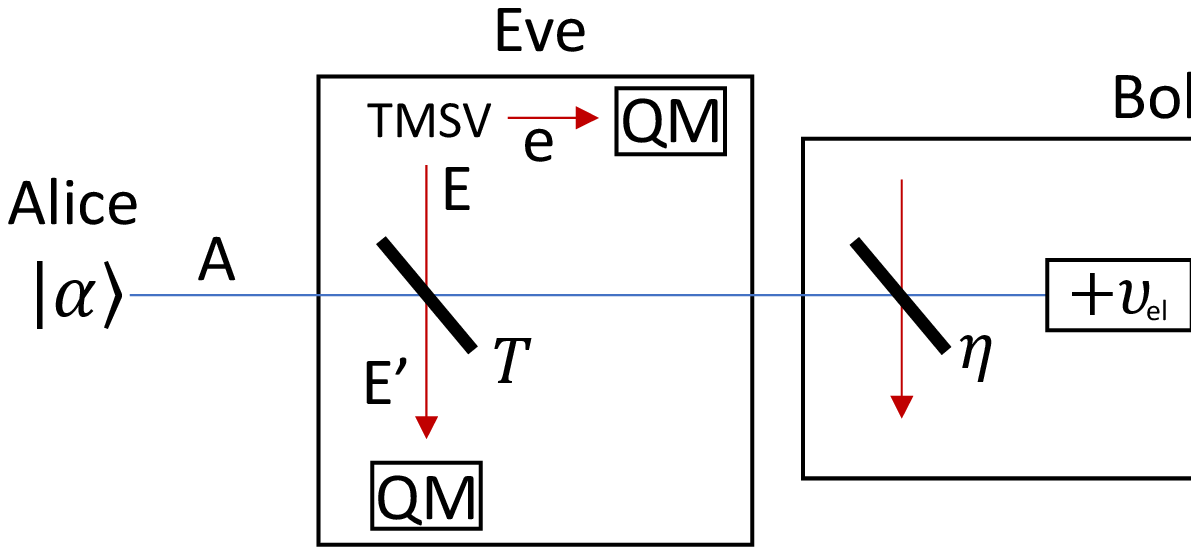}
\end{center}
\par
\vspace{-2.2cm}\caption{Structure of a CV-QKD protocol with (Gaussian-modulated) coherent states considering a
receiver that might have trusted levels of inefficiency and electronic noise. In the middle, the thermal-loss channel
is induced by the collective Gaussian attack of Eve, who uses a beam splitter with transmissivity $T$ and
a TMSV state with variance $\omega$. Eve's outputs are stored in quantum memories (QM) for a later quantum measurement, whose optimal performance is bounded by the Holevo information.}%
\label{fig:setupPIC}%
\end{figure}

\section{Coherent-state protocol and security analysis\label{sec:coherentanalysis}}

\subsection{General description and mutual information\label{sec:mi}}

Assume that Alice prepares a bosonic mode $A$ in
a coherent state $\left\vert \alpha\right\rangle $ whose amplitude is
Gaussian-modulated (see Fig.~\ref{fig:setupPIC} for a schematic). In other words,
we may write $\alpha=(q+ip)/2$, where $x=q,p$ is the mean value of the generic
quadrature $\hat{x}=\hat{q},\hat{p}$, which is randomly chosen according to a
zero-mean Gaussian distribution $\mathcal{N}(0,\sigma_{x}^{2})$ with variance
$\sigma_{x}^{2}:=\mu-1\geq0$, i.e., according to the Gaussian probability density%
\begin{equation}
p_\mathcal{N}(x)=(\sqrt{2\pi}\sigma_{x})^{-1}\exp[-x^{2}/(2\sigma_{x}^{2})].
\end{equation}
Note that we have $q=2\operatorname{Re}\alpha$ and $p=2\operatorname{Im}%
\alpha$ (in fact, we use the notation of Ref.~\cite[Sec.~II]{Gaussian_rev}, where
$[\hat{q},\hat{p}]=2i$ and the vacuum state has noise-variance equal to $1$). Then, $\sigma_{x}^{2}$ represents the modulation variance while $\mu$ is the total signal variance (including the vacuum noise).

The coherent state is sent through an optical fiber of length $L$, which can
be modelled as a thermal-loss channel with transmissivity $T=10^{-\frac{A
L}{10}}$ (e.g., $A=0.2~$dB/km) and thermal noise $\omega=2\bar{n}+1$, where
$\bar{n}$ is the thermal number associated with an environmental mode $E$. The
process can equivalently be represented by a beam splitter with transmissivity
$T$ mixing Alice's mode $A$ with mode $E$, which is in a thermal state with
$\bar{n}$ mean photons. The environmental thermal state can be purified in a
two-mode squeezed vacuum state (TMSV) $\Phi_{eE}$, i.e., a Gaussian state for
modes $e$ and $E$, with zero mean and covariance matrix (CM)~\cite{Gaussian_rev}
\begin{equation}
\mathbf{V}_{eE}(\omega)=\left(
\begin{array}
[c]{cc}%
\omega\mathbf{I} & \sqrt{\omega^{2}-1}\mathbf{Z}\\
\sqrt{\omega^{2}-1}\mathbf{Z} & \omega\mathbf{I}%
\end{array}
\right)  ,
\end{equation}
where $\mathbf{I}:=\mathrm{diag}(1,1)$ and $\mathbf{Z}:=\mathrm{diag}(1,-1)$.
This dilation based on a beamsplitter and a TMSV state is known as an
\textquotedblleft entangling cloner\textquotedblright\ attack\ and represents a realistic collective Gaussian
attack~\cite{collectiveG} (see Fig.~\ref{fig:setupPIC}). Recall that the class of collective Gaussian attacks is optimal for Gaussian-modulated CV-QKD protocols~\cite{revQKD}.

At the other end of the channel, Bob measures the incoming state using a
homodyne detector with quantum efficiency $\eta$ and electronic
noise $\upsilon_{\text{el}}$ (we may also include local coupling losses in parameter $\eta$). The detector is randomly switched between the two
quadratures. Let us
denote by $\hat{y}$ the generic quadrature of Bob's mode $B$ just before the
(ideal) homodyne detector. Then, the outcome $y$ of the  detector satisfies the
input-output formula
\begin{equation}
y=\sqrt{T\eta}x+z, \label{IOrel}
\end{equation}
where $z$ is (or approximately is) a Gaussian noise variable with zero mean and variance
\begin{equation}
\sigma_{z}^{2}=1+\upsilon_{\text{el}}+\eta T\xi, \label{noiseE}
\end{equation}
with $\upsilon_{\text{el}}$ being the electronic noise of the setup and $\xi$
is channel's excess noise, defined by
\begin{equation}
\xi:=\frac{1-T}{T}(\omega-1).\label{excessnoise}
\end{equation}

Here it is important to make two observations. The first consideration is about the general treatment of the detector
at Bob's side, where we assume the potential presence of trusted levels of quantum efficiency
and electronic noise. In the worst-case scenario, these levels can be put equal to zero and
assume that these contributions are implicitly part of the channel transmissivity and excess noise.
In other words, one has $T\eta \rightarrow T$ in Eq.~(\ref{IOrel}), while $\upsilon_{\text{el}}$
becomes part of $\xi$ in Eq.~(\ref{noiseE}), so that $\xi \rightarrow \xi+ \upsilon_{\text{el}}/T$ in Eq.~(\ref{excessnoise}).
The second point is that there might be other imperfections in Alice's and Bob's setups
that are not mitigated or controllable by the parties (e.g., modulation and phase noise).
These imperfections are automatically included in the channel loss and noise via
the general relations of Eqs.~(\ref{IOrel}) and~(\ref{noiseE}). Furthemore, the extra noise
contributions can be considered to be Gaussian in the worst-case scenario, resorting
to the optimality of Gaussian attacks in CV-QKD~\cite{revQKD}.

Assuming the general scenario in Fig.~\ref{fig:setupPIC}, let us compute Alice and Bob's mutual information. From Eq.~(\ref{IOrel}), we
calculate the variance of $y$, which is equal to%
\begin{align}
\sigma_{y}^{2}  &  =T\eta\sigma_{x}^{2}+\sigma_{z}^{2}\nonumber\\
&  =T\eta(\mu-1+\xi)+1+\upsilon_{\text{el}}.
\end{align}
For the conditional variance, we compute
\begin{equation}
\sigma_{y|x}^{2}=\sigma_{y}^{2}(\mu=1)=\eta T\xi+1+\upsilon_{\text{el}}.
\end{equation}
The mutual information associated with the CVs $x$ and $y$ is given
by the difference between the differential entropy $h(y)$ of $y$ and the conditional entropy $h(y|x)$, i.e.,
\begin{align} \label{ABmutual_sigma}
& I(x:y)=h(y)-h(y|x)=\frac{1}{2}\log_{2}\left(  \frac{\sigma_{y}^{2}}{\sigma_{y|x}^{2}
}\right) \\
&  =\frac{1}{2}\log_{2}\left[  1+\frac{\mu-1}{\xi+(1+\upsilon_{\text{el}%
})/T\eta}\right]   \label{ABmutual}\\%
&= \frac{1}{2}\log_{2}\left[  1+\text{SNR}
\right].\label{eq:mutual SNR}
\end{align}
The mutual information has the same value no matter if it is computed
in direct reconciliation (DR), where Bob infers Alice's variable, or reverse reconciliation (RR), where Alice infers Bob's
variable.

As we can see from the expression above, the mutual information contains the
signal-to-noise (SNR) term
\begin{equation}
    \text{SNR}=(\mu-1)/\mathcal{X},
\end{equation}
where
\begin{equation}
\mathcal{X}=\xi+(1+\upsilon_{\text{el}})/T\eta
\end{equation}
is known as equivalent noise. The joint Gaussian distribution of the two variables $x$ and $y$
has zero mean and the following CM
\begin{equation}
   \boldsymbol{\Sigma}_{xy}=\begin{pmatrix}
  \sigma_{x}^{2}&\rho \sigma_{x} \sigma_{y}\\
  \rho \sigma_{x} \sigma_{y}&\sigma_{y}^{2}
  \end{pmatrix}
\end{equation}
where $\rho=\mathbb{E}(x y)/\sigma_{x} \sigma_{y}$ is a correlation parameter (here $\mathbb{E}$ denotes the expected value). From the classical formula $I(x:y)=-(1/2)\log_2(1-\rho^2)$~\cite{CoverThomas},
we see that
\begin{equation}\label{eq:corr}
\rho=\sqrt{\frac{\mathrm{SNR}}{1+\mathrm{SNR}}}.
\end{equation}

It is important to stress that, in order to asymptotically achieve the maximum number of shared bits
$I(x:y)$ per channel use in RR, Bob needs to send $h(y|x)$ bits through the public channel according to Slepian-Wolf coding~\cite{slepian,Wyner}. In practice, Alice and Bob will perform a suboptimal procedure of EC/reconciliation revealing more information $\text{leak}_\text{EC}\geq h(y|x)$. To account for this, one assumes that only a portion $\beta I(x:y)$ of the mutual information can be achieved by using
the reconciliation parameter $\beta\in [0,1)$. This is crucial parameter that depends on various technical quantities as we will see later.

In the ideal case of perfect reconciliation ($\beta=1$), the mutual information
between the parties depends monotonically on the SNR. However, in a realistic case where the reconciliation is not efficient ($\beta<1$),
the extra leaked information $(1-\beta)I(x:y)$  also depends on the SNR. Therefore, by trying to
balance the terms $\beta I(x:y)$ and $\text{leak}_\text{EC}=h(y|x)+(1-\beta)I(x:y)$, one finds an
optimal value for the SNR and, therefore, an optimal value for the total signal variance $\mu$. In practice, this optimal working point
can be precisely computed only after $T$ and $\xi$ are estimated through the procedure of PE (which is discussed later in Sec.~\ref{sec:PE}).
However, a rough estimate of this optimal value can be obtained by an educated guess of the channel parameters (e.g., the approximate transmissivity could be guessed from the length of the fiber and the expected standard loss rate).

\subsection{Asymptotic key rate}

Let us compute the secret key rate that the parties would be able to achieve
if they could use the quantum communication channel an infinite number of
times. Besides $\beta I(x:y)$, we need to calculate Eve's Holevo information
$\chi(E:y)$ on Bob's outcome $y$. This is in fact the maximum information that
Eve can steal under the assumption of collective attacks, where she perturbs
the channel in a independent and identical way, while storing all her outputs
in a quantum memory (to be optimally measured at the end of the protocol). It
is important to stress that, for any processing done by Bob $y\rightarrow
y^{\prime}$, we have $\chi(E:y^{\prime})\leq\chi(E:y)$ so that the latter
value can always be taken as an upper bound for the actual eavesdropping performance.

Let us introduce the entanglement-based
representation of the protocol, where Alice's input ensemble of coherent
states is generated on mode $A$ by heterodyning mode $A^{\prime}$ of a TMSV
$\Phi_{A^{\prime}A}$\ with variance $\mu$. Note that this representation
is not strictly necessary in our analysis (which may be carried over in prepare
and measure completely), but we adopt it anyway for completeness,
so as to give the total state with all the correlations between Alice, Bob and Eve.

The output modes $A^{\prime}$ and
$B$ shared by the parties will be in a zero-mean Gaussian state $\rho
_{A^{\prime}B}$ with CM~\cite{Gaussian_rev}
\begin{equation}
\mathbf{V}_{A^{\prime}B}=\left(
\begin{array}
[c]{cc}%
\mu\mathbf{I} & c\mathbf{Z}\\
c\mathbf{Z} & b\mathbf{I}%
\end{array}
\right)  ,
\end{equation}
where we have set%
\begin{align}
c  &  :=\sqrt{T\eta(\mu^{2}-1)},\\
b  &  :=T\eta(\mu+\xi)+1-T\eta+\upsilon_{\text{el}}.
\end{align}
Then, the global output state $\rho_{A^{\prime}BeE^{\prime}}$ of Alice, Bob
and Eve is zero-mean Gaussian with CM%
\begin{equation}
\mathbf{V}_{A^{\prime}BeE^{\prime}}=\left(
\begin{array}
[c]{cccc}%
\mu\mathbf{I} & c\mathbf{Z} & \mathbf{0} & \zeta\mathbf{Z}\\
c\mathbf{Z} & b\mathbf{I} & \gamma\mathbf{Z} & \theta\mathbf{I}\\
\mathbf{0} & \gamma\mathbf{Z} & \omega\mathbf{I} & \psi\mathbf{Z}\\
\zeta\mathbf{Z} & \theta\mathbf{I} & \psi\mathbf{Z} & \phi\mathbf{I}
\end{array}
\right)  ,
\end{equation}
where $\mathbf{0}$ is the $2\times2$ zero matrix and we set
\begin{align}
\gamma &  :=\sqrt{\eta(1-T)(\omega^{2}-1)},\\
\zeta &  :=-\sqrt{(1-T)(\mu^{2}-1)},\\
\theta &  :=\sqrt{\eta T(1-T)}(\omega-\mu),\\
\psi &  :=\sqrt{T(\omega^{2}-1)},\\
\phi &  :=T\omega+(1-T)\mu.
\end{align}

To compute the Holevo bound, we need to derive the von Neumann entropies
$S(\rho_{eE^{\prime}})$ and $S(\rho_{eE^{\prime}|y})$ which can be computed
from the symplectic spectra of the reduced CM $\mathbf{V}_{eE^{\prime}}$ and
the conditional CM $\mathbf{V}_{eE^{\prime}|y}$. Setting
\begin{equation}
\mathbf{C}=\left(
\begin{array}
[c]{cc}
\gamma\mathbf{Z} & \theta\mathbf{I}
\end{array}
\right),~\boldsymbol{\Pi}:=\left(
\begin{array}
[c]{cc}
1 & 0\\
0 & 0
\end{array}
\right)  ,
\end{equation}
we may use the pseudo-inverse to compute~\cite{Gaussian_rev}
\begin{align}
\mathbf{V}_{eE^{\prime}|y}  &  =\mathbf{V}_{eE^{\prime}}-\mathbf{C}
^{T}[\boldsymbol{\Pi}(b\mathbf{I})\boldsymbol{\Pi}]^{-1}\mathbf{C}\\
&  =\mathbf{V}_{eE^{\prime}}-b^{-1}\mathbf{C}^{T}\boldsymbol{\Pi}\mathbf{C}\\
&  =\left(
\begin{array}
[c]{cc}%
\omega\mathbf{I} & \psi\mathbf{Z}\\
\psi\mathbf{Z} & \phi\mathbf{I}%
\end{array}
\right)  -b^{-1}\left(
\begin{array}
[c]{cc}%
\gamma^{2}\boldsymbol{\Pi} & \gamma\theta\boldsymbol{\Pi}\\
\gamma\theta\boldsymbol{\Pi} & \theta^{2}\boldsymbol{\Pi}%
\end{array}
\right).
\end{align}
Since both $\mathbf{V}_{eE^{\prime}}$ and $\mathbf{V}_{eE^{\prime}|y}$ are
two-mode CMs, it is easy to compute their symplectic spectra, that we denote
by $\{\nu_{\pm}\}$ and $\{\tilde{\nu}_{\pm}\}$, respectively. Their general
analytical expressions are too cumbersome to show here 
unless we take the limit $\mu\gg1$. For instance, we have $\nu
_{+}\rightarrow(1-T)\mu$ and $\nu_{-}\rightarrow\omega$.

Finally, we may write the Holevo bound as%
\begin{align}
\chi(E  &  :y)=S(\rho_{eE^{\prime}})-S(\rho_{eE^{\prime}|y})\nonumber\\
&  =h(\nu_{+})+h(\nu_{-})-h(\tilde{\nu}_{+})-h(\tilde{\nu}_{-}),
\label{HolevoB}%
\end{align}
where
\begin{equation}
h(\nu):=\frac{\nu+1}{2}\log_{2}\frac{\nu+1}{2}-\frac{\nu-1}{2}\log_{2}%
\frac{\nu-1}{2}.
\end{equation}
The asymptotic key rate is given by%
\begin{align}
R_{\text{asy}}  &  =\beta I(x:y)-\chi(E:y)\\
&  =R(\beta,\mu,\eta,\upsilon_{\text{el}},T,\xi)\label{eq:asymptotic}.
\end{align}
Suppose that $\eta$ and $\upsilon_{\text{el}}$ are known and the parties have preliminary
estimates of $T$ and $\xi$. Then, for a target $\beta$, Alice can compute
an optimal value $\mu_{\mathrm{opt}}$ for the signal variance $\mu$ by optimizing the asymptotic rate in Eq.~(\ref{eq:asymptotic}).

\subsection{Parameter estimation\label{PEsection}\label{sec:PE}}

In a realistic implementation of the protocol, the parties use the quantum channel a finite number $N$ of times. The first consequence of this
finite-size scenario is that their knowledge of the channel parameters $T$ and
$\xi$ is not perfect. Thus, once the quantum communication is over, Alice and Bob
declare $m$ random instances $\{x_{i}\}$ and $\{y_{i}\}$ of their local
variables $x$ and $y$.
From these instances, they build the maximum-likelihood
estimators $\widehat{T}$ of the  transmissivity $T$\ and  $\widehat{\Xi}$ of the excess noise variance $\Xi:=\eta T \xi$, for which they also exploit their knowledge of the trusted levels of detector/setup efficiency $\eta$ and electronic noise $\upsilon_{\text{el}}$.

Following Ref.~\cite{UsenkoFNSZ}, we write
\begin{equation}
    \widehat{T}=\frac{1}{\eta\sigma_x^4} \left(\widehat{C}_{xy}\right)^2
\end{equation}
where
\begin{equation}
    \widehat{C}_{xy}=\frac{1}{m}\sum_{i=1}^m x_iy_i,
\end{equation}
is the estimator for the covariance $C_{xy}=\sqrt{\eta T}\sigma_x^2$ between $x$ and $y$. This covariance is normally distributed with following mean and variance
\begin{align}
\mathbb{E}( \widehat{C}_{xy})&=\sqrt{\eta T}\sigma_x^2=C_{xy},\label{eq:meanC}\\
\text{Var}( \widehat{C}_{xy})&=\frac{1}{m}\left(2\eta T  \left(\sigma_x^2\right)^2+\sigma_x^2 \sigma^2_z\right):=V_{C_{xy}}\label{eq:varC}.
\end{align}

Then one can express $\widehat{T}$ as a (scaled) non-central chi-squared variable
\begin{equation}
\widehat{T}=\frac{V_{C_{xy}}}{\eta^2\left(\sigma_x^2\right)^2} \left(\frac{\widehat{C}_{xy}}{\sqrt{V_{C_{xy}}}}\right)^2,
\end{equation}
since $\widehat{C}_{xy}/\sqrt{V_{C_{xy}}}$ follows a standard normal distribution.
The mean and variance of $\widehat{T}$ is given by the associated noncentral chi-squared parameters $k=1$ and $\lambda=\frac{C^2_{xy}}{V_{C_{xy}}}$. Therefore, we obtain
\begin{equation}
\mathbb{E}(\widehat{T})=\frac{V_{C_{xy}}}{\eta^2\left(\sigma_x^2\right)^2}\left(1+\frac{C^2_{xy}}{V_{C_{xy}}}\right),
\end{equation}
and
\begin{equation}
\text{Var}(\widehat{T})=\frac{2V^2_{C_{xy}}}{\eta^4\left(\sigma_x^2\right)^4}\left(1+2\frac{C^2_{xy}}{V_{C_{xy}}}\right).
\end{equation}
Using Eqs.~(\ref{eq:meanC}) and~(\ref{eq:varC}), and keeping only the significant terms with respect to $1/m$, we have
\begin{equation}\label{eq:sigmaT}
\mathbb{E}(\widehat{T})=T,~\text{Var}(\widehat{T})=\frac{4}{m}T^2\left(2+\frac{\sigma_z^2}{\eta T \sigma^2_x}\right):=\sigma_T^2.
\end{equation}

The estimator for the noise variance $\sigma^2_z$ is given by
\begin{equation}
\widehat{\sigma^2_z}=\frac{1}{m}\sum_{i=1}^m\left(y_i-\sqrt{\eta \widehat{T}}x_i\right)^2.
\end{equation}
Assuming $\hat{T}\approx T$ and rescaling by $1/\sigma_z$ the term inside the brackets in the relation above, we get a standard normal distribution for the variable $\frac{z_i}{\sigma_z}$ with $z_i=y_i-\sqrt{\eta T}x_i$. Thus, we obtain
\begin{equation}
\widehat{\sigma^2_z}=\frac{\sigma_z^2}{m}\sum_{i=1}^m\left(\frac{z_i}{\sigma_z}\right)^2,
\end{equation}
and observe that this is a (scaled)  chi-squared variable. From the associated chi-squared parameters $k=m$ (and $\lambda=0$), we calculate the following mean and variance
\begin{align}
\mathbb{E}(\widehat{\sigma_z^2})&=\sigma_z^2,\\
\text{Var}(\widehat{\sigma_z^2})&=\frac{2\left(\sigma_z^2\right)^2}{m}.
\end{align}
Then, from the formula
\begin{equation}
\sigma^2_z=1+v_\text{el}+\Xi,
\end{equation}
we have that
\begin{align}
&\widehat{\Xi}=\widehat{\sigma_z^2}-\upsilon_\text{el}-1,\\
&\mathbb{E}(\widehat{\Xi})=\Xi,\\
&\text{Var}(\widehat{\Xi})=\frac{2\left(\sigma_z^2\right)^2}{m}:=\sigma_\Xi^2\label{eq:sigmaXi}.
\end{align}

For $m$ sufficiently large, and up to an error probability
$\varepsilon_{\text{PE}}$, the channel parameters fall in the intervals%
\begin{align}
T &  \in\lbrack \widehat{T}-w\sigma_{\widehat{T}},\widehat{T}+w \sigma_{\widehat{T}} ],\\
\Xi &  \in\lbrack\widehat{\Xi}-w\sigma_{\widehat{\Xi}},\widehat{\Xi}+w\sigma_{\widehat{\Xi}}],
\end{align}
where $\sigma_{\widehat{T}}$, $\sigma_{\widehat{\Xi}}$ are given by the Eq.~(\ref{eq:sigmaT}) and~(\ref{eq:sigmaXi}), where we replace the actual values $T$ and $\sigma_z^2$ with their corresponding estimators. Note that $w$ is expressed in terms of $\varepsilon_{\text{PE}}$ via the inverse error function, i.e.,
\begin{equation}
    w=\sqrt{2}\operatorname{erf}^{-1}(1-\varepsilon_{\text{PE}}). \label{errordisplayed}
\end{equation}
The worst-case estimators will be given by
\begin{equation}\label{eq:wcvalues}
T_m=\widehat{T}-w \sigma_{\widehat{T}},~\Xi_m=\widehat{\Xi}+w \sigma_{\widehat{\Xi}}
\end{equation}
so we have $T\geq T_{m}$ and $\Xi\leq {\Xi}_{m}$ up to an error
$\varepsilon_{\text{PE}}$ (see Appendix~\ref{PEapp} for other details on the derivation of $T_{m}$).

In the next step, Alice and Bob compute an overestimation of Eve's Holevo bound in terms
of $T_{m}$ and $\Xi_{m}$, so that they may write the modified rate%
\begin{equation}
R_{m}:=\beta I(x:y)-\left.  \chi(E:y)\right\vert _{T_{m},\Xi_{m}}. \label{eq:Rm}
\end{equation}
Accounting for the number of signals sacrificed for PE, the
actual rate in terms of bits per channel use is given by the rescaling
\begin{equation}
R_{m}\rightarrow\frac{n}{N}R_{m},
\end{equation}
where $n=N-m$ are the instances for key generation.

Note that, from the estimators $\widehat{T}$ and $\widehat{\Xi}$, the parties may
compute an estimator for the SNR, i.e.,
\begin{equation}\label{eq:est_SNR}
\widehat{\mathrm{SNR}}=\frac{(\mu-1)\eta\widehat{T}}{1+\upsilon_{\text{el}}%
+\widehat{\Xi}}.
 \end{equation}
Therefore, in a more practical implementation, the rate in Eq.~(\ref{eq:Rm}) is replaced by the following expression
\begin{align}
  &  R_{m} =\beta \left. I(x:y)\right\vert_{\widehat{T},\widehat{\Xi}} -\left. \chi(E:y)\right\vert_{T_{m},\Xi_{m}}, \label{Rmrate} \\
 &   \left. I(x:y)\right\vert_{\widehat{T},\widehat{\Xi}} = \frac{1}{2}\log_{2}\left(  1+\widehat{\mathrm{SNR}}\right). \label{Iestimated}
\end{align}
For a fixed $\beta$, the parties could potentially optimize the SNR over the signal variance $\mu$. In fact, before the quantum communication,
they may use rough estimates about transmissivity and excess noise to compute value $\mu_\text{opt}(\beta)$
that maximizes the asymptotic key rate.

In a practical implementation,
the data generated by the QKD\ protocol is sliced in $n_{\text{bks}}\gg1$
blocks, each block being associated with the quantum communication of $N$ points (modes).
Assuming that the channel is sufficiently stable over time, the statistics (estimators and worst-case values) can be
computed over
\begin{equation}
M:=mn_{\text{bks}}\gg m
\end{equation}
random instances, so that all the estimators (i.e, $\widehat{T}$, $\widehat{\Xi}$, $\widehat{\mathrm{SNR}}$, $T_{m}$, and $\Xi_{m}$) are computed over $M$ points and we replace $R_{m}\rightarrow R_{M}$ in
Eq.~(\ref{Rmrate}), i.e., we consider
\begin{equation}
 R_{M} =\beta \left. I(x:y)\right\vert_{\widehat{T},\widehat{\Xi}} -\left. \chi(E:y)\right\vert_{T_{M},\Xi_{M}}, \label{Rmrate2}
\end{equation}

This also means that we consider an average of $m=M/n_{\text{bks}}$ points for
PE in each block, and an average number of $n=N-M/n_{\text{bks}}$ key generation points from
each block to be processed in the step of EC. Because $N$ is typically large, the variations around the averages can be considered to be negligible,
which means that we may assume $m$ and $n$ to be the actual values for each block.

Finally note that, if the channel instead varies over a
timescale comparable to the block size $N$ (which is a condition that may occur in free-space quantum communications~\cite{Panosfading,freeSPACE}), then we may need to perform
PE independently for each block. In this case, one would have
a different rate for each block, so that the final key rate will be given by an
average. However, for ground-based fiber-implementations, the channel is typically
stable over long times, which is the condition assumed here.

\subsection{Error correction\label{sec:EC}}
Once PE has been done, the parties process their remaining $n$ pairs (key generation points) in a procedure of EC.
Here we combine elements from various works~\cite{Mario,Pacher_LDPC,MacKayNeal,Mackay2,univershal_hashing}.
The procedure can be broken down in steps of normalization, discretization, splitting, LDPC encoding/decoding, and EC verification.

\subsubsection{Normalization} In each block of size $N$, Alice and Bob have $n$ pairs $\{x_i,y_i\}$ of
their variables $x$ and $y$ that are related by Eq.~(\ref{IOrel}) and can be used for key generation.
As a first step, Alice and Bob normalize their variables 
by dividing them by the respective standard deviations, i.e.,~\cite{Mario}
\begin{equation}
x\rightarrow X:=x/\sigma_{x},~~y\rightarrow Y:=y/\sigma_{y},   \label{normVARs}
\end{equation}
so $X$ and $Y$ have the following CM
\begin{equation}
  \boldsymbol{\Sigma}_{XY}=\begin{pmatrix}
  1&\rho\\
  \rho&1
  \end{pmatrix}.
\end{equation}
Variables $X$ and $Y$ follow a standard normal bivariate
distribution with correlation $\rho=\mathbb{E}(X Y)$, which is connected to the SNR by Eq.~(\ref{eq:corr}), where the latter is approximated by Eq.~(\ref{eq:est_SNR}) in a practical scenario.

Under conditions of stability for the quantum communication channel, the normalization in Eq.~(\ref{normVARs}) is performed over the entire
record of $n n_{\mathrm{bks}}$ key generation points. Using the computed standard deviations $\sigma_{x}$ and $\sigma_{y}$, we then build the strings $X^{n}=x^{n}/\sigma_{x}$ and
$Y^{n}=y^{n}/\sigma_{y}$ for each block, starting from the corresponding $n$ key generation points $x^{n} = \{x_i\}$ and $y^{n} = \{y_i\}$.

\subsubsection{Discretization\label{sec:disc}} Bob discretizes his normalized variable $Y$ in a $p$-ary variable  $K$ with generic value $\kappa \in \{0,\dots,2^p-1\}$ being an element of a Galois field $\mathcal{GF}(2^p)$ (see Appendix~\ref{ap:GLFD}).
This is achieved as follows. As a first step, he sets a cut-off $\alpha$ such that $|Y|\leq \alpha$
occurs with negligible probability, which is approximately true for
$\alpha \geq 3$. Then, Bob chooses the size $\delta=2 \alpha 2^{-p}$  of the intervals (bins) $[a_\kappa,b_\kappa)$ of his lattice,
whose border points are given by~\cite{Pacher_LDPC}
\begin{equation}\label{eq:left border}
 a_\kappa=\begin{cases}
 -\infty~~~~~~~~\text{for}~~\kappa=0,\\\\
 - \alpha + \kappa \delta~~~\text{for}~~\kappa>0,
 \end{cases}
\end{equation}
and
\begin{equation}\label{eq:right border}
 b_\kappa=\begin{cases}
 -\alpha + (\kappa+1) \delta~~~\text{for}~\kappa<2^p-1,\\\\
 \infty~~~~~~~~~~~~~~~~~~~\text{for}~\kappa=2^p-1.
 \end{cases}
\end{equation}
Finally, for any value of $Y \in [a_\kappa,b_\kappa)$, Bob takes $K$ equal to $\kappa$. Thus, for $n$ points, the normalized string $Y^{n}$ is transformed into a string of discrete values $K^{n}$. Note that this discretization technique is very basic. Indeed one could increase the performance by adopting bins of different sizes depending on the estimated SNR.

\subsubsection{Splitting} 
\begin{figure}[t]
\vspace{0.2cm}
\includegraphics[width=0.45\textwidth]{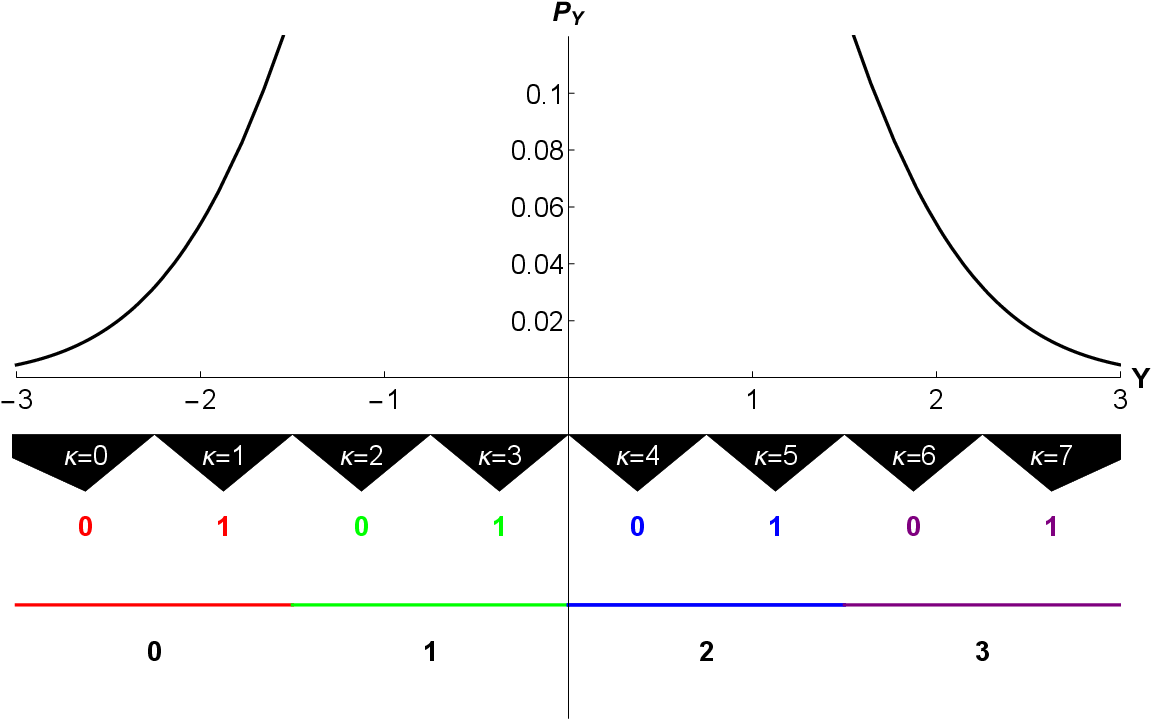}
\caption{Discretization and splitting with $\alpha=3$, $p=3$ and $q=2$. The variable $Y$ follows a normal distribution $P_Y$ so that the probability of $|Y|>3$ is assumed to be negligible. Variable $Y$ and the bins defined in Eqs.~(\ref{eq:left border}) and~(\ref{eq:right border}) identify a discrete variable $K$ with values $\kappa=0,\dots,7$ (black triangles). During the splitting stage, each bin can be described by two numbers: $\overline{\kappa}=0\dots,3$ associated with $q=2$, and $\underline{\kappa}=0,1$ associated with $d=p-q=1$. We see that $2^d$ bins belong to each super bin $\overline{\kappa}$ (colored intervals). Bob uses the parity check matrix of a non-binary LDPC code to encode the information $\overline{\kappa}$ related to the super bin while the position $\underline{\kappa}$ of the bin inside the super bin is broadcast through the public channel.}%
\label{fig:splitting_fig}%
\end{figure}

Bob sets an integer value for $q<p$ and computes $d=p-q$. Then, he splits his discretized variable in two parts $K=(\overline{K},\underline{K})$,
where the top variable $\overline{K}$ is $q$-ary and the bottom variable $\underline{K}$ is $d$-ary. Their values are defined by splitting the generic value $\kappa$ in the following two parts
\begin{equation}
\overline{\kappa}=\frac{\kappa-(\kappa\mod2^d)}{2^d},~~\underline{\kappa}=(\kappa\mod2^d). \label{topbottombits}
\end{equation}
In other words, we have
\begin{equation}
\kappa=\overline{\kappa} 2^d + \underline{\kappa}. \label{topbottombits2}
\end{equation}
With the top variable $\overline{K}$, Bob creates $2^q$ super bins with each super bin containing $2^d$ bins
associated with the bottom variable $\underline{K}$. See also Fig.~\ref{fig:splitting_fig}.

Repeating this for $n$ points provides a string of values $\overline{K}^{n}$ for the super bins and another string for the relative
bin-positions $\underline{K}^{n}$. The most significant string $\overline{K}^{n}$ is locally processed by an LDPC code (more details below),
while the least significant string $\underline{K}^{n}$ is side information that is revealed through the public channel.

\subsubsection{LDPC encoding and decoding\label{LDPCencdec}} 
Bob constructs an $l \times n$ parity check matrix $\mathbf{H}$ with $q$-ary entries from $\mathcal{GF}(2^q)$ according to Ref.~\cite{MacKayNeal}. This matrix is applied to the top string $\overline{K}^{n}$ to derive the $l$-long syndrome $K^{l}_\text{\text{sd}}=\mathbf{H}\overline{K}^{n}$, where the matrix-vector product is defined in $\mathcal{GF}(2^q)$. The syndrome is sent to Alice
together with the direct communication of the bottom string $\underline{K}^{n}$. The parity check matrix is associated with an LDPC code~\cite{Mackay2}, which may encode $k=n-l$ source symbols into $n$ output symbols, so that it has rate
\begin{equation}\label{eq:LDPC RATE}
R_\text{code}:= k/n=1-R_\text{syn},~R_\text{syn}:=l/n.
\end{equation}
What value of $R_\text{code}$ to use and, therefore, what matrix $\mathbf{H}$ to build is explained afterwards in this subsection.

From the knowledge of the syndrome $K^{l}_\text{\text{sd}}$, Bob's bottom string $\underline{K}^{n}$ and her local string $X^{n}$, Alice decodes Bob's top string $\overline{K}^{n}$. This is
done via an iterative belief propagation algorithm~\cite{Mackay2} where, in every iteration, she updates a codeword likelihood function (see also Appendix~\ref{ap:LDPC decoding}).
The initial likelihood function before any iteration comes from the a priori probabilities
\begin{equation}\label{eq:a-priori prob}
 P({\overline{K}|X,\underline{K}})=\frac{P({\overline{K},\underline{K}|X})}{\sum_{\overline{K}} P({\overline{K},\underline{K}|X})},
\end{equation}
where
\begin{align}\label{eq:conditional prob}
&P({\overline{K},\underline{K}|X})=P({K|X})\\&=
\frac{1}{2}\text{erf}\left(\frac{b_{\kappa}-\rho x/\sigma_{x} }{\sqrt{2(1-\rho^2)}}\right)-\frac{1}{2}\text{erf}\left(\frac{a_{\kappa}-\rho x/\sigma_{x}}{\sqrt{2(1-\rho^2)}}\right). \label{initialPROB}
\end{align}
At any iteration $\le \text{iter}_\text{max}$ Alice finds the argument that maximizes the likelihood function. If the syndrome of this argument is equal to $K^{l}_\text{\text{sd}}$, then the argument forms her guess $\widehat{K}^{n}$ of Bob's top string $\overline{K}^{n}$. However, if this syndrome matching test (SMT) is not satisfied within the maximum number of iterations $\text{iter}_\text{max}$, then the block is discarded. The possibility of failure in the SMT reduces the total number of input blocks from $n_{\mathrm{bks}}$ to $n_{\mathrm{SMT}}= p_{\mathrm{SMT}} n_{\mathrm{bks}}$ where $p_{\mathrm{SMT}}$ is the probability of successful matching within the established $\text{iter}_\text{max}$ iterations.

Let us compute the LDPC rate $R_{\text{code}}$ to be used. First notice how Alice and Bob's mutual information decreases as
a consequence of the classical data processing inequality~\cite{CoverThomas} applied
to the procedure. We have
 \begin{align}\label{eq:data_processing}
  I(x:y)&\geq I(X:Y) \\
 &\geq I(X:K)=H(K)-H(K|X)\\
&\geq H(K)-\text{leak}_\text{EC}, \label{eq:data_processing2}
 \end{align}
where $\text{leak}_\text{EC} \ge H(K|X)$ comes from the Wolf-Slepian limit~\cite{slepian,Wyner} 
and $H(K)$ is the Shannon entropy of $K$, which can be computed over the entire record of $n_\text{ent}=n n_{\mathrm{bks}}$ key generation points (under stability conditions).
The parties empirically estimate the entropy $H(K)$. To do so, they build the MLE
\begin{equation}
\widehat{H}(K)=-\sum_{\kappa=0}^{2^p-1}f_{\kappa}\log_2 f_{\kappa},
\end{equation}
where $f_{\kappa}=n_{\kappa}/n_\text{ent}$ stands for the frequencies of the symbols $\kappa$, defined as the ratio between the number of times $n_{\kappa}$ that a symbol appears in the string over the string's length $n_\text{ent}$.
For this estimator, it is easy to show that~\cite{kontogiannis}
\begin{equation}\label{Sentropy bound}
H(K)\geq \widehat{H}(K)-\delta_\text{ent},
\end{equation}
with penalty
\begin{equation}
    \delta_\text{ent}=\log_2(n_\text{ent})\sqrt{\frac{2\log(2/\varepsilon_\text{ent})}{n_\text{ent}}}.\label{entpen}
\end{equation}
This bound is valid up to an error probability $\varepsilon_\text{ent}$.

In Eq.~(\ref{eq:data_processing2}), the leakage $\text{leak}_\text{EC}$ is upper-bounded by the equivalent number
of bits per use that are broadcast after the LDPC encoding in each block, i.e.
\begin{equation}
    \text{leak}_\text{EC} \le d+R_\text{syn}q.
\end{equation}
Therefore, combining the two previous bounds, we may write
\begin{equation}
I(x:y) \ge \beta I(x:y):=\widehat{H}(K)-\delta_\text{ent}+R_\text{code} q-p. \label{codeEQ1first}
\end{equation}

Note that in a practical implementation Alice and Bob do not access $I(x:y)$, but rather $I(x:y) \vert_{\widehat{T},\widehat{\Xi}}$ from Eq.~(\ref{Iestimated}). Therefore, considering this modification in Eq.~(\ref{codeEQ1first}), we more precisely write
\begin{equation}
\beta I(x:y) \vert_{\widehat{T},\widehat{\Xi}} =\widehat{H}(K)-\delta_\text{ent}+R_\text{code} q-p, \label{codeEQ1}
\end{equation}
so the rate in Eq.~(\ref{Rmrate2}) becomes
\begin{align}
 & R_{M}^{\mathrm{EC}} =\widehat{H}(K)-\delta_\text{ent}+R_\text{code} q-p -\left. \chi(E:y)\right\vert_{T_{M},\Xi_{M}}. \label{rateOK} 
\end{align}

From Eqs.~(\ref{Iestimated}) and~(\ref{codeEQ1}), we see that the LDPC code must be chosen to have rate
\begin{align}
 R_\text{code} &\simeq q^{-1} \left[ \frac{\beta}{2}\log_{2}\left(  1+\widehat{\mathrm{SNR}}\right) + p - \widehat{H}(K)+\delta_\text{ent} \right]  \label{RcodeFORMULA0}
 \end{align}
for some estimated SNR and key entropy. A value of the reconciliation efficiency $\beta$ is acceptable only if
we can choose parameters $\alpha \geq 3$ and $q$~\cite{noteI}, such that $R_\text{code} \le 1$.
Once $R_\text{code}$ is known, the sparse parity check matrix $\mathbf{H}$ of the LDPC code can be constructed following Ref.~\cite{MacKayNeal}.

\subsubsection{Verification}
An important final step in the EC procedure is the verification of the $n_{\mathrm{SMT}}$ error-corrected blocks that have successfully passed the SMT. For each of these blocks, the parties possess two $n$-long $q$-ary strings with identical syndromes, i.e., Bob's top string $\overline{K}^{n}$ and Alice's guess $\widehat{K}^{n}$. The parties convert their strings into a binary representation, $\overline{K}^{n}_{\mathrm{bin}}$ and $\widehat{K}^{n}_{\mathrm{bin}}$, so that each of them is $qn$ bit long. In the next step, Alice and Bob compute $t$-bit long hashes of their converted binary strings following Ref.~\cite{univershal_hashing} (universal hash functions are used because of their resistance to collisions). In particular, they set $t = \lceil -\log_2\varepsilon_\text{cor}\rceil$, where $\varepsilon_\text{cor}$ is known as `correctness'.

Then, Bob discloses his hash to Alice, who compares it with hers. If the hashes are identical, the verification stage is deemed successful. The two strings $\overline{K}^{n}_{\mathrm{bin}}$ and $\widehat{K}^{n}_{\mathrm{bin}}$ are identical up to a small error probability $2^{-t} \le \varepsilon_\text{cor}$. In such a case, the associated bottom string $\underline{K}^{n}$ held by both parties is converted by both parties into binary $\underline{K}^{n}_{\mathrm{bin}}$ and appended to the respective strings, i.e.,  $\overline{K}^{n}_{\mathrm{bin}} \rightarrow \overline{K}^{n}_{\mathrm{bin}} \underline{K}^{n}_{\mathrm{bin}}$ and  $\widehat{K}^{n}_{\mathrm{bin}} \rightarrow \widehat{K}^{n}_{\mathrm{bin}} \underline{K}^{n}_{\mathrm{bin}}$. Such binary concatenations are promoted to the next step of PA.

By contrast, if the hashes are different, then the two postprocessed strings are discarded (together with the public bottom string). Therefore, associated with the hash verification test, we have a probability of success that we denote by $p_{\mathrm{ver}}$. This is implicitly connected with $\varepsilon_\mathrm{cor}$. In fact, if we decrease $\varepsilon_\mathrm{cor}$, we increase the length of the hashes to verify, meaning that we increase the probability of spotting an uncorrected error in the strings, leading to a reduction of the success probability $p_{\mathrm{ver}}$.

Thus, combining the possible failures in the syndrome and hash tests, we have a total probability of success for EC, which is given by
\begin{equation}
    p_{\text{EC}} = p_{\mathrm{SMT}} p_{\mathrm{ver}} := 1-p_{\mathrm{FER}},
\end{equation}
where $p_{\mathrm{FER}}$ is known as `frame error rate' (FER).

Note that the effective value of $p_{\text{EC}}$ depends not only on
$\varepsilon_{\text{cor}}$ but also on the specific choice for the LDPC code. In particular,
in the presence of a very noisy channel, using a high-rate
LDPC code (equivalent to a high value of $\beta$) implies low correction
performances and, therefore, a low value for $p_{\text{EC}}$ (SMT and/or hash test tend to fail).
By contrast, the use of a low-rate EC code (low value of $\beta$) implies a high value for
$p_{\text{EC}}$ (good performance for both SMT and hash test).
Thus, there is an implicit trade-off between $p_{\text{EC}}$
and $\beta$. For fixed values of $\varepsilon_{\text{cor}}$ and $\beta$, the
value of $p_{\text{EC}}$ (success of the EC test)\ still depends on the
channel noise, so that it needs to be carefully calculated from the
experimental/simulation points.

\subsection{Privacy amplification and composably-secure finite-size key}
The final step is that of PA, after which the secret key is generated. Starting from the original $n_{\text{bks}}$ blocks (each being $N$-size), the parties derive $p_{\text{EC}}n_{\text{bks}}$ successfully error-corrected binary strings
\begin{equation}
S :=\overline{K}^{n}_{\mathrm{bin}} \underline{K}^{n}_{\mathrm{bin}} \simeq  \widehat{S}:= \widehat{K}^{n}_{\mathrm{bin}} \underline{K}^{n}_{\mathrm{bin}},
\end{equation}
each string containing $n p$ bits. 
In this final step, all the surviving error-corrected strings are compressed into shorter
strings that are decoupled from Eve (up to a small error probability that we discuss below). This compression step can be
implemented on each sequence individually, or globally on a
concatenation of sequences. Here, the latter approach is adopted and is certainly valid under conditions of channel stability.

Concatenating their local $p_{\text{EC}}n_{\text{bks}}$ error-corrected strings,
Alice and Bob construct two long binary sequences $\mathbf{S} \simeq \widehat{\mathbf{S}}$,
each having $\tilde{n}:= p_{\text{EC}}n_{\text{bks}} n p$ bits. Each of these sequences will be
compressed to a final secret key of $r:= p_{\text{EC}} n_{\text{bks}} n\tilde{R}$ bits, where $\tilde{R}$
is determined by the composable key rate (see below). The compression is achieved via universal hashing
by applying a Toeplitz matrix $\mathbf{T}_{r,\tilde{n}}$ which is calculated efficiently with the use of FFT (more details in Appendix~\ref{ap:toeplitz_sk}). Thus, from their sequences, Alice and Bob
finally retrieve the secret key
\begin{equation}
\mathbf{K}= \mathbf{T}_{r,\tilde{n}} \mathbf{S} \simeq   \mathbf{T}_{r,\tilde{n}} \widehat{\mathbf{S}}.  \label{finalKEYexpression}
\end{equation}

An important parameter to consider for the step of PA is the `secrecy' $\varepsilon_{\text{sec}}$,
which bounds the distance between the final key and an ideal key from which Eve
is completely decoupled. Technically, one further decomposes\ $\varepsilon
_{\text{sec}}=\varepsilon_{\text{s}}+\varepsilon_{\text{h}}$, where
$\varepsilon_{\text{s}}$ is a smoothing parameter $\varepsilon_{\text{s}}$ and
$\varepsilon_{\text{h}}$ is a hashing parameter. These PA epsilon parameters are combined
with the parameter from EC.

Let us call $\tilde{\rho}^{n}$ the classical-classical-quantum state shared by Alice, Bob and Eve after EC. We may write \begin{equation}\label{distance wc}
  p_\text{EC}  D(\tilde{\rho}^{n}, \rho_\text{id}) \leq \varepsilon:= \varepsilon_\text{sec} +\varepsilon_\text{cor},
\end{equation}
where $\varepsilon$ is the epsilon security of the protocol, $D$ is the trace distance and $\rho_\text{id}$ is the output of an ideal protocol where Eve is completely decoupled from Bob, with Alice's and Bob's keys being exactly the same ~\cite[App.~G]{freeSPACE}.

Accounting for the estimation of the channel parameters [cf.~Eq.~(\ref{eq:wcvalues})] and the key entropy [cf.~Eq.~(\ref{Sentropy bound})] is equivalent to replacing  $\tilde{\rho}^{n}$ with a worst-case state $\tilde{\rho}_{\text{wc}}^{n}$ in the computation of the key rate. However, there is a small probability $\varepsilon^\prime_{\text{PE}}$ that we have a different state $\rho_{\text{bad}}$ violating one or more of the tail bounds associated with the worst-case estimators. This means that, on average, we have the state \begin{equation}
\rho_\text{PE}=(1-\varepsilon^\prime_{\text{PE}})\tilde{\rho}_{\text{wc}}^{n}+\varepsilon^\prime_{\text{PE}}\rho_\text{bad}.
\end{equation}

Because we impose $p_\text{EC}  D(\tilde{\rho}_{\text{wc}}^{n}, \rho_\text{id}) \leq \varepsilon$ and the previous equation implies $ D(\tilde{\rho}_{\text{wc}}^{n}, \rho_\text{PE}) \leq \varepsilon^\prime_{\text{PE}}$, the triangle inequality provides $ p_\text{EC} D( \rho_\text{PE}, \rho_\text{id})\leq  \varepsilon + p_\text{EC} \varepsilon^\prime_{\text{PE}}$, meaning that the imperfect parameter estimation adds an extra term $p_\text{EC} \varepsilon^\prime_{\text{PE}}$ to the epsilon security of the protocol. In other words, we have that the protocol is secure up to re-defining  $\varepsilon \rightarrow \varepsilon + p_\text{EC} \varepsilon^\prime_{\text{PE}}$. Note that we have
\begin{align}
\varepsilon^\prime_{\text{PE}} &=(1-2\varepsilon_\text{PE})\varepsilon_\text{ent}+(1-\varepsilon_\text{ent})2\varepsilon_\text{PE}+2\varepsilon_\text{PE}\varepsilon_\text{ent} \\ &\simeq2\varepsilon_\text{PE}+\varepsilon_\text{ent},
\end{align}
so we can write
\begin{align}
\varepsilon &= \varepsilon_\text{sec} +\varepsilon_\text{cor} + p_\text{EC} \varepsilon^\prime_{\text{PE}}
\\ &\simeq \varepsilon_\text{s} +\varepsilon_\text{h}+\varepsilon_\text{cor}+p_\text{EC}(2\varepsilon_\text{PE}+\varepsilon_\text{ent}).\label{overallEPS}
\end{align}
A typical choice is to set
$\varepsilon_{\text{s}}=\varepsilon_{\text{h}}=\varepsilon_{\text{cor}%
}=\varepsilon_\text{PE}=\varepsilon_{\text{ent}}=2^{-32}\simeq2.3\times10^{-10}$, so that
$\varepsilon\lesssim 4 \times 10^{-9}$ for any $p_{\text{EC}}$. 

For success probability $p_{\text{EC}}$ and $\varepsilon$-security against
collective (Gaussian) attacks, the secret key rate of the protocol (bits per channel use) takes the form~\cite[Eq.~(105)]{freeSPACE}%
\begin{equation}
R=\frac{np_{\text{EC}}}{N}\tilde{R},~\tilde{R}:=\left(R_{M}^{\mathrm{EC}}-\frac{\Delta_{\text{AEP}}}{\sqrt{n}}+\frac{\Theta}{n}\right), \label{rateCOMPO}
\end{equation}
where $R_{M}^{\mathrm{EC}}$ is given in Eq.~(\ref{rateOK}) and the extra terms are equal to the following~\cite{QKDlevels}
\begin{align}
&  \Delta_{\text{AEP}}:=4\log_{2}\left(2^{p/2} +2\right)  \sqrt{\log
_{2}\left(  \frac{18}{p_{\text{EC}}^{2}\varepsilon_{\text{s}}^{4}}\right)
},\label{deltaAEPPP}\\
&  \Theta:=\log_{2}[p_{\text{EC}}(1-\varepsilon_{\text{s}}^{2}/3)]+2\log
_{2}\sqrt{2}\varepsilon_{\text{h}}. \label{bigOMEGA}
\end{align}
Note that the discretization bits $p$ appear in $\Delta_{\text{AEP}}$~\cite{noteRATE}

The practical secret key rate in Eq.~(\ref{rateCOMPO}) can be compared with a corresponding theoretical rate
\begin{equation}
R_{\mathrm{theo}}=\frac{n \tilde{p}_{\text{EC}}}{N}\left(R^{*}_{M}-\frac{\tilde{\Delta}_{\text{AEP}}}{\sqrt{n}}+\frac{\tilde{\Theta}}{n}\right), \label{rateTHEO}
\end{equation}
where $\tilde{p}_{\text{EC}}$ is guessed, with $\tilde{\Delta}_{\text{AEP}}$ and $\tilde{\Theta}$ being computed on that guess. Then, we have
\begin{equation}
R_{M}^{*} = \tilde{\beta} \left. I(x:y)\right\vert_{T,\Xi} -\left. \chi(E:y)\right\vert_{T^{*}_{M},\Xi^{*}_{M}},
\end{equation}
where $\tilde{\beta}$ is also guessed, and the various estimators are approximated by their mean values, so that $\widehat{T} \simeq T$, $\widehat{\Xi} \simeq \Xi=\eta T \xi$ and we have set
\begin{align}
T^{*}_{M} &=T-w\sigma_T, \\
\Xi^{*}_M &=\Xi+w \sigma_\Xi,
\end{align}
with $w$ depending on $\varepsilon_{\text{PE}}$ as in Eq.~(\ref{errordisplayed}).

\section{Protocol simulation and data processing\label{sec:processing}}
Here we sequentially go over the steps of the protocol and its postprocessing, as they need to be implemented in a numerical simulation or an actual experimental demonstration. We provide more technical details and finally present the pseudocode of the entire procedure.

\subsection{Main parameters}
We start by discussing the main parameters related to the physical setup,
communication channel and protocol. Some of these parameters are taken as input,
while others need to be estimated by the parties, so that they represent
output values of the simulation.

\begin{description}
\item[Setup:] Main parameters are Alice's total signal variance $\mu$, Bob's trusted levels of local efficiency $\eta
$, and electronic noise $\upsilon_{\text{el}}$. These are all input values.

\item[Channel:] Main parameters are the effective transmissivity $T$, and excess noise $\xi$. These are input values
to our simulation, which are used to create the input-output relation of Eq.~(\ref{IOrel}). In an experimental
implementation, these values are generally unknown and Eq.~(\ref{IOrel}) comes from the experimental data.

\item[Protocol:] Main parameters are the number of blocks $n_{\text{bks}}$, the size of each block $N$,
the total number $M$ of instances for PE, the various epsilon parameters $\varepsilon_{\text{s,h,...}}$
and the $p$-bit discretization, so the alphabet size is $D=2^p$. These are all input values. Output values are the estimators $\widehat{T}$,
$\widehat{\Xi}$, $T_{M}$, $\Xi_M$, the EC probability of success $p_{\text{EC}}$, the reconciliation parameter $\beta$, the final rate $R$ and key sting $\mathbf{K}$.
\end{description}

In Tables~\ref{table1} and~\ref{table2}, we summarize the main input and output parameters. In Table~\ref{table3}, we schematically show the formulas
for other related parameters.

\begin{table}[h!]
\begin{tabular}
[c]{|l|l|}
\hline
parameter & description \\\hline
$L$ & Channel length (km) \\
$A$ & Attenuation rate (dB/km) \\
$\xi$ & Excess noise  \\
$\eta$ & Detector/Setup efficiency  \\
$\upsilon_{\text{el}}$ & Electronic noise  \\
$\beta$ & (Target) reconciliation efficiency\\
$n_{\text{bks}}$ & Number of blocks  \\
$N$ & Block size  \\
$M$ & Number of PE\ runs  \\
$p$ & Discretization bits   \\
$q$ & Most significant (top) bits \\
$\alpha$ & Phase-space cut-off \\
$\text{iter}_\text{max}$ & Max number of EC iterations  \\
$\varepsilon_{\text{\text{PE}, s, h, \text{corr}}}$ & Epsilon parameters  \\
$\mu$ & Total signal variance \\
\hline
\end{tabular}\caption{Main input parameters}\label{table1}
\end{table}

\begin{table}[h!]
\begin{tabular}
[c]{|l|l|}
\hline
parameter & description \\\hline
$\mu_\text{opt}$ & Optimal signal variance  \\
$R_\text{asy}$ & Asymptotic key rate  \\
$\widehat{T}$, $\widehat{\Xi}$, $T_{M}$, $\Xi_{M}$ & Channel estimators  \\
$\widehat{\text{SNR}}$ & Estimated SNR
\\ $\widehat{H}(K)$ & Key entropy estimator \\
$R_\text{code}$ & Code rate   \\
$p_{\text{EC}}$ & EC success probability\\
$\text{fnd}_\text{rnd}$ & EC syndrome matching round  \\
$r$ & Final key length  \\
$R$ & Composable key rate  \\
$\mathbf{K}$ & Final key  \\
$\varepsilon$ & $\varepsilon$-security  \\
\hline
\end{tabular}\caption{Main output parameters}\label{table2}
\end{table}

\begin{table}[h]
\begin{tabular}
[c]{|l|l|l|}%
\hline
parameter & description & formula \\\hline
$T$ & Channel transmissivity & $10^{{-AL} / {10}}$ \\
$\sigma^2_z$ & Noise variance & $1 + v_\text{el} + \eta  T  \xi$ \\
$\Xi$ & Excess noise variance & $\eta  T \xi$ \\
$\omega$ & Thermal noise & $\frac{T\xi - T + 1}{1 - T}$ \\
$\mathcal{X}$ & Equivalent noise & $\xi +\frac{ 1 + v_\text{el}}{T\eta}$ \\
$\text{SNR}$ & Signal-to-noise ratio & $(\mu - 1)/\mathcal{X}$ \\
$m$ & PE instances per block & $M / n_\text{bks}$ \\
$n$ & Key generation points per block & $N - m$ \\
$\text{FER}$ & Frame error rate & $1-p_{\mathrm{EC}}$   \\
$\text{GF}$& Number of the $\mathcal{GF}$ elements & $2^q$\\
$\delta$ & Lattice step in phase space & $\frac{\alpha}{2^{p-1}}$ \\
$d$ & Least significant (bottom) bits & $p - q$ \\
$t$ & Verification hash output length & $\lceil-\log_2\varepsilon_{\text{cor}}\rceil$ \\
$\rho$ & Correlation coefficient & $\sqrt{ \frac{\widehat{\text{SNR}}}{1 + \widehat{\text{SNR}}}}$ \\

$\delta_\text{ent}$ & Entropy penalty & 
See Eq.~(\ref{entpen})

\\

$\tilde n$ & Total bit string length after EC &
$n\;p\;n_{bks}\;p_\text{EC}$\\
\hline
\end{tabular}\caption{Related parameters}\label{table3}
\end{table}

\subsection{Quantum communication}

The process of quantum communication can be simplified in the following two steps before and after the action
of the channel:

\begin{description}
\item[Preparation:] Alice encodes $Nn_{\text{bks}}$ instances $\{x_{i}\}$ of the mean $x$ of the generic quadrature $\hat{x}$, such that $x\sim\mathcal{N}(0,\mu-1)$. In the experimental practice, the two conjugate quadratures are independently encoded in the amplitude of a coherent state, but only one of them will survive after
the procedure of sifting (here implicitly assumed).

\item[Measurement:] After the channel and the projection of (a randomly-switched) homodyne detection, Bob decodes $Nn_{\text{bks}}$ instances
$\{y_{i}\}$ of $y=\sqrt{T\eta}x+z$, where the noise variable $z\sim\mathcal{N}(0,\sigma_{z}^{2})$
has variance $\sigma_z^{2}$ as in Eq.~(\ref{noiseE}).
\end{description}

As mentioned above, it is implicitly assumed that Alice and Bob perform a sifting stage where Bob
classically communicates to Alice which quadrature he has measured (so that
the other quadrature is discarded).

\subsection{Parameter estimation}

The stage of PE is described by the following steps:

\begin{description}
\item[Random positions:] Alice randomly picks $M$ positions $i\in
\lbrack1,Nn_{\text{bks}}]$, say $\{i_{u}\}_{u=1}^{M}$. On average
$m=M/n_{\text{bks}}$ positions are therefore picked from each block, and $n=N-m$ points
are left for key generation in each block (for large enough blocks,
the spread around these averages is negligible).

\item[Public declaration:] Using a classical channel, Alice communicates the
$M$ pairs $\{i_{u},x_{i_{u}}\}_{u=1}^{M}$ to Bob.

\item[Estimators:] Bob sets a PE error $\varepsilon_{\text{PE}}$. From the
pairs $\{x_{i_{u}},y_{i_{u}}\}_{u=1}^{M}$, he computes the estimators
$\widehat{T}$ and $\widehat{\Xi}$, and the worst-case estimators $T_{M}$ and
$\Xi_{M}$ for the channel parameters (see formulas in Sec.~\ref{PEsection}).

\item[Early termination:] Bob checks the threshold condition $\left.
I(x:y)\right\vert _{\widehat{T},\widehat{\Xi}}>\left.  \chi(E:y)\right\vert
_{T_{M},\Xi_{M}}$, which is computed in terms of the estimators $\widehat{T}$ and $\widehat
{\Xi}$, and worst-case estimators
$T_{M}$ and $\Xi_{M}$ (associated with $\varepsilon_{\text{PE}}$).
If the threshold condition is not satisfied, then the protocol is aborted.
\end{description}

\subsection{Error correction}

The procedure of EC is performed on each block of size $N$
and consists of the following steps:

\begin{description}

\item[Normalization:]
For key generation, Alice and Bob have $n$ pairs $\{x_i,y_i\}$ of
their variables $x$ and $y$ that are related by Eq.~(\ref{IOrel}). As a first step, Alice and Bob normalize their variables $x$ and $y$ according to Eq.~(\ref{normVARs}),
therefore creating $X$ and $Y$.

\item[Discretization:]
Bob sets the cut-off value $\alpha$ and the step $\delta=\alpha 2^{1-p}$ of his lattice, whose generic bin
$Y[a_\kappa,b_\kappa)$ is defined by Eqs.~(\ref{eq:left border}) and~(\ref{eq:right border}). Then, he discretizes
his normalized variable $Y$ into a $p$-ary variable $K$ with generic value $\kappa \in \{0,\dots,2^p-1\}$ with the following rule:
For any value of his variable $Y \in [a_\kappa,b_\kappa)$, Bob takes $K$ equal to $\kappa$.

\item[Splitting:] Bob sets an integer value for $q$ and computes $d=p-q$. From his discretized variable $K$, he creates the top $q$-ary variable $\overline{K}$
and the bottom $d$-ary variable $\underline{K}$, whose generic values $\overline{\kappa}$ and $\underline{\kappa}$ are defined by Eq.~(\ref{topbottombits}).
For $n$ points, he therefore creates a string $\overline{K}^{n}$ which is locally processed via an LDPC code (see below),
and another string $\underline{K}^{n}$ which is revealed through the public channel.

\item[LDPC encoding:] From the SNR estimator $\widehat{\text{SNR}}$, the entropy estimator $\widehat{H}(K)$, and a target reconciliation efficiency $\beta$, the parties use Eq.~(\ref{RcodeFORMULA0}) to derive the rate $R_{\text{code}}$ of the LDPC code. They then build its $l \times n$ parity-check matrix $\mathbf{H}$ with $q$-ary entries from $\mathcal{GF}(2^{q})$, using the procedure of Ref.~\cite{MacKayNeal}, i.e., (i) the column weight $d_v$ (number of nonzero elements in a column) is constant and we set $d_v = 2$; (ii) the row weight $d_c$ adapts to the formula $R_\text{code} = 1 - d_v/d_c$ and is as uniform as possible; and (iii) the overlap (inner product between two columns) is never larger than $1$.
Once $\mathbf{H}$ is constructed, Bob computes the syndrome $K^{l}_\text{\text{sd}}=\mathbf{H}\overline{K}^{n}$, which is sent to Alice together with the bottom string $\underline{K}^{n}$.

\item[LDPC decoding:]From the knowledge of the syndrome $K^{l}_\text{\text{sd}}$, Bob's bottom string $\underline{K}^{n}$ and her local string $X^{n}$, Alice decodes her guess
$\widehat{K}^{n}$ of Bob's top string $\overline{K}^{n}$. This is done via a sum-product algorithm~\cite{Mackay2}, where in each iteration $\text{iter}< \text{iter}_\text{max}$ Alice  updates a suitable likelihood function with initial value given by the a priori probabilities in Eq.~(\ref{initialPROB}). If the syndrome of $\widehat{K}^{n}$ is equal to $K^{l}_\text{\text{sd}}$, then Alice's guess $\widehat{K}^{n}$ of Bob's top string $\overline{K}^{n}$ is promoted to the next verification step. If the syndrome matching test fails for $\text{iter}_\text{max}$ iterations, the block is discarded and the frequency/probability $1-p_{\mathrm{SMT}}$ of this event is registered. For more details of the sum-product algorithm see Appendix~\ref{ap:LDPC decoding}.

\item[Verification:] Each pair of promoted strings $\overline{K}^{n}$ and $\widehat{K}^{n}$ is converted to binary $\overline{K}^{n}_{\mathrm{bin}}$ and $\widehat{K}^{n}_{\mathrm{bin}}$. Over these, the parties compute hashes of $t = \lceil -\log_2\varepsilon_\text{cor}\rceil$ bits. Bob discloses his hash to Alice, who compares it with hers. If the hashes are identical, the parties convert the corresponding bottom string $\underline{K}^{n}$ into binary $\underline{K}^{n}_{\mathrm{bin}}$ and promote the two concatenations
\begin{equation}
S :=\overline{K}^{n}_{\mathrm{bin}} \underline{K}^{n}_{\mathrm{bin}} \simeq  \widehat{S}:= \widehat{K}^{n}_{\mathrm{bin}} \underline{K}^{n}_{\mathrm{bin}} \label{afterEC}
\end{equation}
to the next step of PA. By contrast, if the hashes are different, then $\overline{K}^{n}_{\mathrm{bin}}$ and $\widehat{K}^{n}_{\mathrm{bin}}$ are discarded, together with $\underline{K}^{n}$. The parties compute the frequency/probability of success of the hash verification test $p_{\mathrm{ver}}$ and derive the overall success probability of EC $p_{\mathrm{EC}}= p_{\mathrm{SMT}} p_{\mathrm{ver}} = 1-  p_{\mathrm{FER}}$.
In more detail, the strings $\overline{K}^{n}_{\mathrm{bin}}$ and $\widehat{K}^{n}_{\mathrm{bin}}$ are broken into $Q$-bit strings that are converted to $Q$-ary numbers forming the strings $\overline{K}^{n^\prime}_Q$ and $\widehat{K}^{n^\prime}_Q$ with $n^\prime =n q/Q$ symbols for $Q>q$. In case $n q/Q$ is not an integer, the strings $\overline{K}^{n}_{\mathrm{bin}}$ and $\widehat{K}^{n}_{\mathrm{bin}}$ are padded with $s q$ zeros so that $n^\prime=(n+s) q/Q \in \mathbb{N}$. Bob then derives independent uniform random integers $v_i=1,\dots,2^{Q^*}-1$, where $v_{i}$ is odd, and an integer $u=0,\dots,2^{Q^*}-1$,
for $i=1,\dots,n^\prime$ and ${Q^*}\leq Q+t-1$
with $\varepsilon_\text{cor}\leq2^{-t}$ being the target collision probability. 
After Bob communicates his choice of universal families to Alice, they both hash each of the $Q$-ary numbers and combine the results according to the following formula~\cite{univershal_hashing}
\begin{equation}
 \tilde{h}\left(\mathbf{x}\right)=\left(\sum_{i=1}^{n^\prime}v_{i}x_{i}\right)+u, \label{hashEQ}
\end{equation}
where $\mathbf{x}=\overline{K}^{n^\prime}_Q$ (for Bob) or $\widehat{K}_Q^{n^\prime}$ (for Alice). Summation and multiplication in Eq.~(\ref{hashEQ}) are modulo $2^{Q^*}$. In practice, this is done by discarding the overflow (number of bits  over  $Q^*$) of $\tilde{h}\left(\mathbf{x}\right)$. Then they keep only the first $t$ bits to form the hashes (where typically $t=32$).
\end{description}

\subsection{Privacy amplification}
After EC, Alice and Bob are left with $p_{\text{EC}}n_{\text{bks}}$ successfully error-corrected
binary strings, each of them being represented by Eq.~(\ref{afterEC}) and containing $n p$ bits.
By concatenation, they build two long binary sequences $\mathbf{S} \simeq \widehat{\mathbf{S}}$,
each having $\tilde{n}:= p_{\text{EC}}n_{\text{bks}} n p$ bits. For the chosen level of secrecy
$\varepsilon_{\text{sec}}$, the parties compute the overall epsilon security $\varepsilon$ from Eq.~(\ref{overallEPS}) and the key rate $R=\frac{np_{\text{EC}}}{N}\tilde{R}$ according to Eq.~(\ref{rateCOMPO}).
Finally, the sequences $\mathbf{S} \simeq \widehat{\mathbf{S}}$ are compressed into a final secret key of length $r:= p_{\text{EC}} n_{\text{bks}} n\tilde{R}$ bits by applying a Toeplitz matrix $\mathbf{T}_{r,\tilde{n}}$, so the secret key is given by Eq.~(\ref{finalKEYexpression}).
\bigskip

\subsection{Pseudocode of the procedure}
In Algorithm~1, we present the entire pseudocode of the procedure, whose steps are implemented in Python~\cite{library}.

\begin{algorithm}
	\caption{High-Level Routine Overview}
	\begin{algorithmic}[1]
		\State $L, A, \eta, \xi, v_\text{el}, n_\text{bks}, N, M, \varepsilon_\text{PE}, \varepsilon_s, \varepsilon_h, \varepsilon_{cor}, \beta, \text{iter}_\text{max}, p, q,\alpha \leftarrow$ Input\_Values\_Definition()
		\State $T, \sigma^2_z, \Xi, m, n, t, \mathrm{GF}, \delta, d  \leftarrow \text{Dependent\_Values()}$
		\State Validity\_Checks()
		\If {is\_mu\_optimal}
		\State $\mu_\text{opt} \leftarrow$ Optimal\_Signal\_Variance()
		\Else
		\State $\mu \leftarrow$ user\_input
		\EndIf
		\For{$\text{blk}=1, 2, \dots n_\text{bks}$}
		\State $x[\text{blk}] \leftarrow$ State\_Preparation()
		\State $y[\text{blk}] \leftarrow$ State\_Transmission()
		\State $y[\text{blk}] \leftarrow$ State\_Measurement()
		\State $x[\text{blk}] \leftarrow$ Key\_Sifting()
		\EndFor
		\State $R_\text{asy}, \left.I(x:y)\right\vert _{T,\Xi},  \left.  \chi(E:y)\right\vert_{T,\Xi} \leftarrow$ Rate\_Calculation()
		\State $\{i_{u},x_{i_{u}}\}_{u=1}^{M}, \{i_{u},y_{i_{u}}\}_{u=1}^{M} \leftarrow$Sacrificed\_States\_Selection()
		\State $\hat{T}, \hat{\Xi}, T_{M}, \Xi_{M} \leftarrow$ Parameter\_Estimation()
		\State $R_{M}, \left.I(x:y)\right\vert _{\widehat{T},\widehat{\Xi}}, \left.  \chi(E:y)\right\vert_{T_{M},\Xi_{M}} \leftarrow$ Rate\_Calculation()
		\If {$\left.I(x:y)\right\vert _{\widehat{T},\widehat{\Xi}}\leq\left.  \chi(E:y)\right\vert_{T_{M},\Xi_{M}}$}
		\State Abort\_Protocol()
		\EndIf
		\State $X, Y \leftarrow$ Normalization()
		\State $\widehat{\text{SNR}}, \rho \leftarrow$ Code\_Estimations()
		\For{$\text{blk}=1, 2, \dots n_\text{bks}$}
		\State $K[\text{blk}] \leftarrow$ Discretization()
		\State $\overline{K}[\text{blk}], \underline{K}[\text{blk}] \leftarrow$ Splitting()
		\State $p_{\overline{K}|X,\underline{K}}[\text{blk}] \leftarrow$ A\_Priori\_Probabilities\_Calculation()
		\EndFor
		\State $\widehat{H}$(K),$R_\text{code} \leftarrow$ Code\_Rate\_Calculation()
		\State $\mathbf{H} \leftarrow$ LDPC\_Code\_Generation()
		\For{$\text{blk}=1, 2, \dots n_\text{bks}$}
		    \State $\overline{K}^l_\text{\text{sd}}[\text{blk}] \leftarrow$ Bob\_Syndrome\_Calculation()
		    \State $\widehat{K}^n$[\text{blk}], fnd, $\text{rnd}_\text{fnd}$ $\leftarrow$ Non\_Binary\_Decoding()
		    \State $\widehat{K}^{n}_{\mathrm{bin}}[\text{blk}],\overline{K}^{n}_{\mathrm{bin}}[\text{blk}], \underline{K}^{n}_{\mathrm{bin}}[\text{blk}] \leftarrow$ Bin\_Conversion()
		    \State $\text{hash\_verified[\text{blk}]} \leftarrow$ Verification()
		 \If {is\_$\text{hash\_verified}[\text{blk}]$}
		    \State $\widehat{S}[\text{blk}] \leftarrow$ Concatenate($\widehat{K}^{n}_{\mathrm{bin}}[\text{blk}]$, $\underline{K}^{n}_{\mathrm{bin}}[\text{blk}]$)
 		    \State $S[\text{blk}] \leftarrow$ Concatenate($\overline{K}^{n}_{\mathrm{bin}}[\text{blk}]$, $ \underline{K}^{n}_{\mathrm{bin}}[\text{blk}]$)
		    \EndIf
		\EndFor
		\State $p_\text{EC}, \text{FER} \leftarrow$ Frame\_Error\_Rate\_Calculation()
		\State $R, r, \Tilde{n}, \varepsilon \leftarrow$ Composable\_Rate\_Calculation()
		\If {$R > 0$}
		    \For{$\text{blk}=1, 2, \dots p_\text{EC}n_\text{bks}$}
		    \State $\hat{\mathbf{S}} \leftarrow$ $\text{Append}( \widehat{S}[\text{blk}])$
		   \State $\mathbf{S} \leftarrow$ $ \text{Append}( S[\text{blk}])$
            \EndFor
		   \State  $\mathbf{K} \leftarrow$ Privacy\_Amplification()
		\EndIf
		\State Information\_Logging()
	\end{algorithmic}
\end{algorithm}

\section{Simulation results\label{sec:simulations}}

We are particularly interested in short-range high-rate implementations of CV-QKD, over distances of around $5$~km in standard optical fiber. Even in this regime of relatively high SNR, to get a positive value for the composable secret key rate, we need to consider a block size $N$ of the order of at least $10^5$. The choice of the reconciliation efficiency $\beta$ is also important, as a positive rate cannot be achieved when the value of $\beta$ is too low.

\begin{figure}[t]
    \vspace{-0.5cm}
    \centering
    \includegraphics[width=0.45\textwidth]{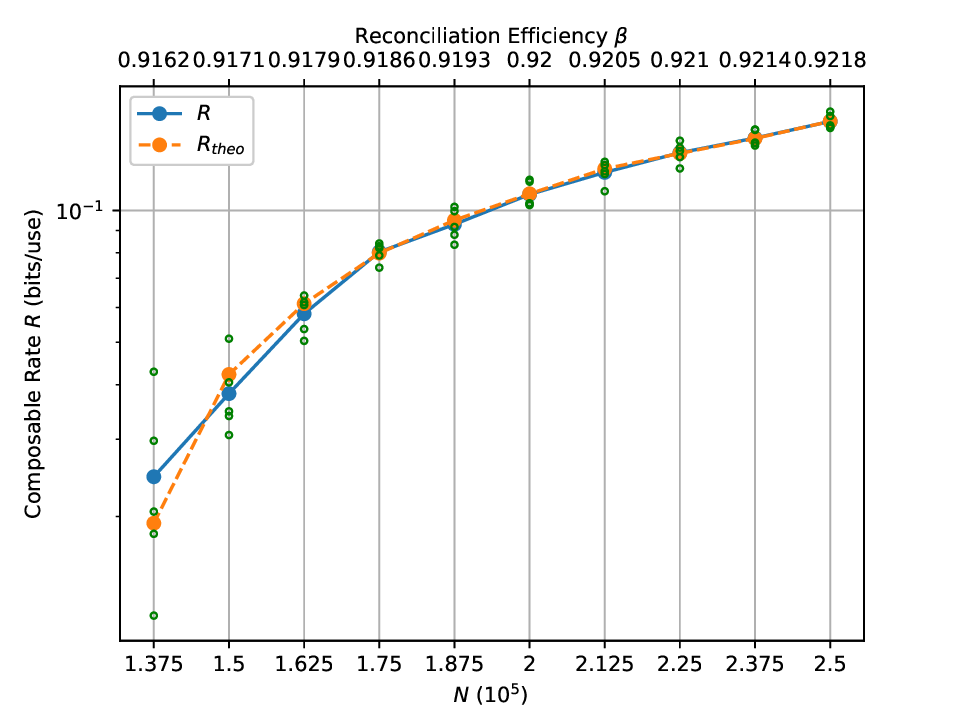}
    \vspace{-0.2cm}
    \caption{Composable secret key rate $R$ (bits/use) versus the block size $N$ for $\text{SNR}=12$. We compare the rate of Eq.~(\ref{rateCOMPO}) from five simulations (green points) and their average (blue line) with the theoretical rate of Eq.~(\ref{rateTHEO}) (orange line), where the theoretical guesses for $\tilde{\beta}$ and $\tilde{p}_{\mathrm{EC}}$ are chosen compatibly with the simulations. For every simulation, $\tilde{p}_{\mathrm{EC}}=p_{\mathrm{EC}}$ has been set. All simulations have achieved $p_\text{EC} \geq 0.95$.  The step of $N$ is $12500$. The values of the reconciliation efficiency $\beta$ are shown on the top axis and are chosen so as to produce $R_\text{code} \approx 0.875$. See Table~\ref{res_param1} for the list of input parameters used in the simulations.}
    \label{fig:R_vs_N}
\end{figure}

\begin{figure}[t]
    \vspace{-0.5cm}
    \centering
    \includegraphics[width=0.45\textwidth]{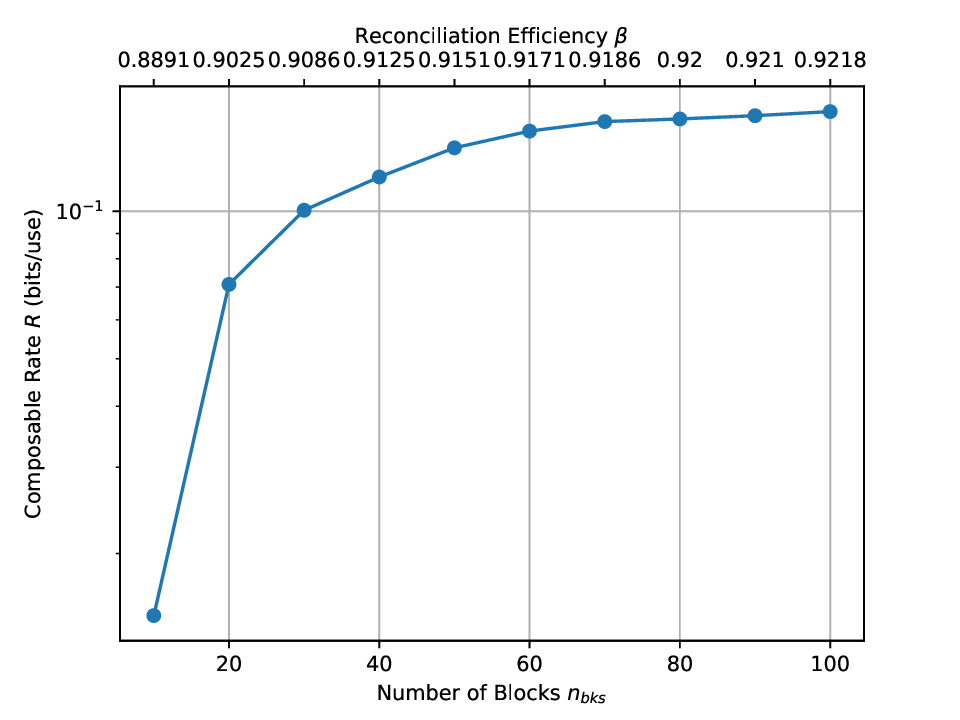}
    \vspace{-0.2cm}
    \caption{Composable secret key rate $R$ (bits/use) versus the number of blocks $n_\text{bks}$ for $\text{SNR}=12$. The step of $n_\text{bks}$ is $10$. The individual block size is fixed and equal to $N=2.5 \times 10^5$. Every point represents the average value of $R$, which is obtained after 5 simulations. All simulations have achieved $p_\text{EC} \geq 0.95$. The values of the reconciliation efficiency $\beta$ are shown on the top axis and are chosen so as to produce $R_\text{code} \approx 0.875$. See Table~\ref{res_param1} for the list of input parameters used in the simulations.}
    \label{fig:R_vs_n_bks}
\end{figure}

Sample parameters for a positive $R$ are given in Table~\ref{res_param1}. Alice's signal variance $\mu$ is chosen to achieve a target high value of SNR (e.g., $\text{SNR} = 12$ in Figs.~\ref{fig:R_vs_N} and~\ref{fig:R_vs_n_bks}). As we see from Fig.~\ref{fig:R_vs_N}, positive values for the composable secret key rate are indeed achievable for block sizes with $N>10^{5}$ and, as expected, the key rate grows as the block size increases, while all of the simulations attained $p_\text{EC} \geq 0.95$. The numerical values of the rate can be considered to be high, since a key rate of $10^{-1}$ bits/use corresponds to $500$~kbits/sec with a relatively slow clock of $5$~MHz. Fig.~\ref{fig:R_vs_n_bks} implies that high rates can be achieved even with fewer total states $Nn_\text{bks}$, when a large block size, e.g. $N=250000$, is fixed and the number of blocks $n_\text{blk}$ varies instead. Note that having an adequately large block size is much more beneficial in obtaining a positive $R$ than having more blocks of smaller sizes. Having fewer total states also achieves faster performance, as seen in Fig. \ref{benchmarks}.

\begin{figure}[t]
    \vspace{-0.5cm}
    \centering
    \includegraphics[width=0.45\textwidth]{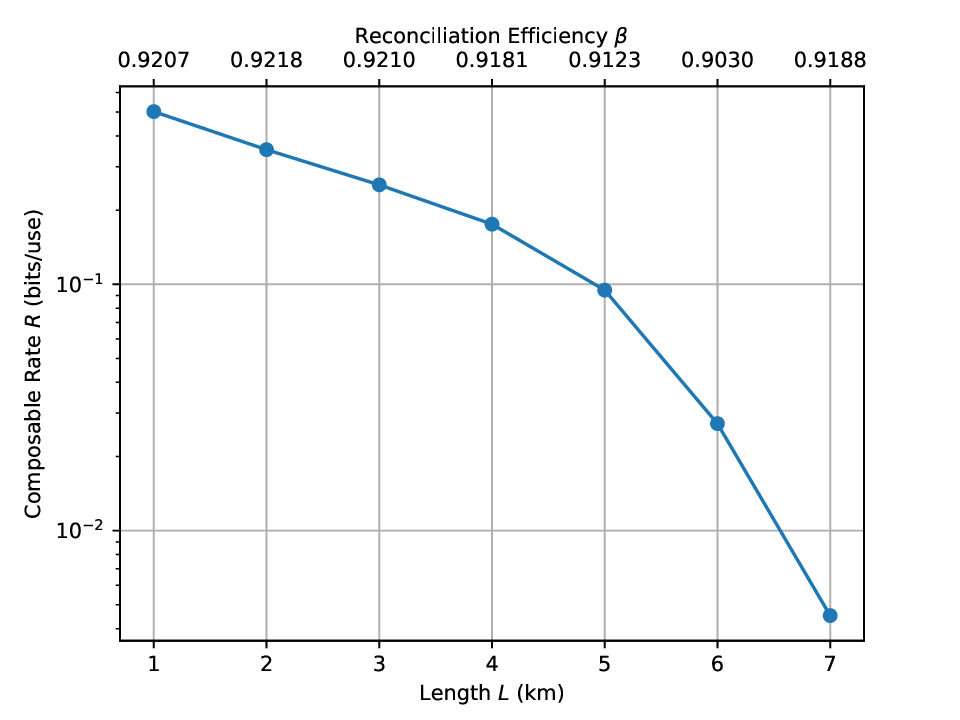}
    \vspace{-0.2cm}
    \caption{Composable secret key rate $R$ (bits/use) versus the channel length $L$ (km). Here, we use $N=2 \times 10^{5}$. Every point represents the average value of $R$, which is obtained after 5 simulations. All simulations have achieved $p_\text{EC} \geq 0.95$. The values of the reconciliation efficiency $\beta$ are shown on the top axis. Other parameters are taken as in Table~\ref{res_param1}.}
    \label{fig:R_vs_L}
\end{figure}

In Fig.~\ref{fig:R_vs_L}, we also show the behavior of the composable secret key rate versus distance $L$ expressed in km of standard optical fiber. We adopt the input parameters specified in Table~\ref{res_param1} and we use blocks of size $N=2 \times 10^{5}$, with reconciliation efficiency $\beta$ taking values from $90.25\%$ to $92.17\%$. As we can see from the figure, high rates (around $0.5$ bits/use) can be achieved at short distances ($L=1$~km), while a distance of $L=7$~km can yield a rate of about $0.004$ bits/use.

\begin{figure}[t]
    \vspace{-0.2cm}
    \centering
    \includegraphics[width=0.45\textwidth]{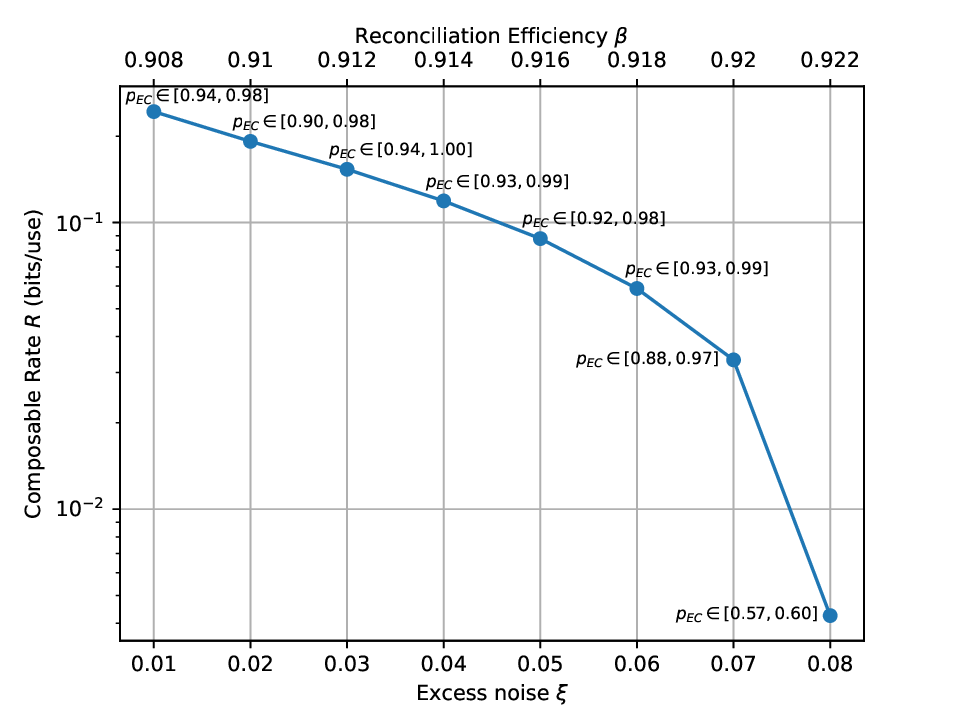}
    \vspace{-0.2cm}
    \caption{Composable secret key rate $R$ (bits/use) versus the excess noise $\xi$. Every point represents the average value of $R$, which is obtained after 5 simulations. The minimum and maximum values achieved for $p_\text{EC}$ fall within the interval displayed next to each point. The values of the reconciliation efficiency $\beta$ are shown on the top axis and are chosen so as to produce $R_\text{code} \approx 0.913$. Other parameters are taken as in Table~\ref{res_param1}.}
    \label{fig:R_vs_xi}
\end{figure}

In Fig.~\ref{fig:R_vs_xi}, we analyze the robustness of the protocol with respect to the amount of untrusted excess noise in the quantum communication channel (even though this parameter may also include any other imperfection coming from the experimental setup). As we can see from the figure, positive key rates are achievable for relatively high values of the excess noise ($\xi = 0.08$).

\begin{table*}[t]
\vspace{0.1cm}
\begin{tabular}
[c]{|c|c|c|c|c|c|c|}\hline
Parameter & Value (Fig.~\ref{fig:R_vs_N}) & Value (Fig.~\ref{fig:R_vs_n_bks})
& Value (Fig.~\ref{fig:R_vs_L}) & Value (Fig.~\ref{fig:R_vs_xi}) & Value (Fig.~\ref{fig:FER_vs_SNR}) & Value (Figs.~\ref{fig:R_vs_SNR}-\ref{fig:avg_rnd_vs_SNR})\\ \hline
$L$ & $5$ & $5$ & variable & $4$ & $5$ & $5$ \\
$A$ & $0.2$ & $0.2$ & $0.2$ & $0.2$ & $0.2$ & $0.2$ \\
$\xi$ & $0.01$ & $0.01$ & $0.01$ & variable & $0.01$ & $0.01$ \\
$\eta$ & $0.8$ & $0.8$ & $0.8$ & $0.85$ & $0.8$ & $0.8$ \\
$\upsilon_{\text{el}}$ & $0.1$ & $0.1$ & $0.1$ & $0.05$ & $0.1$ & $0.1$\\
$n_{\text{bks}}$ & $100$ & variable & $100$ & $100$ & $100$ & $100$\\
$N$ & $1.375$-$2.5\times10^{5}$ & $2.5\times 10^{5}$ & $2\times10^{5}$ & $2.5\times10^{5}$ & $2\times10^{5}$ & $2.5\times 10^{5}$ \\
$M$ & $0.1n_{\text{bks}}N$ & $0.1n_{\text{bks}}N$ & $0.1n_{\text{bks}}N$ & $0.1n_{\text{bks}}N$ & $0.1n_{\text{bks}}N$ & $0.1n_{\text{bks}}N$ \\
$p$ & $7$ & $7$ & $7$ & $7$ & $7$ & variable \\
$q$ & $4$ & $4$ & $4$ & $4$ & $4$ & $4$ \\
$\alpha$ & $7$ & $7$ & $7$ & $7$ & $7$ & $7$ \\
$\text{iter}_{\text{max}}$ & $100$ & $100$ & $150$ & $200$ & $150$ & $150$\\
$\varepsilon_{\text{\text{PE}, s, h, \text{...}}}$ & $2^{-32}$ & $2^{-32}$ & $2^{-32}$ & $2^{-32}$ & $2^{-32}$ & $2^{-32}$\\
$\mu$ & $\approx21.89$ & $\approx21.89$ & $20$ & $25$ & variable & variable \\\hline
\end{tabular}
\caption{The input parameters for the simulations.}\label{res_param1}
\end{table*}


\begin{figure}[t]
    \vspace{-0.1cm}
    \centering
    \includegraphics[width=0.45\textwidth]{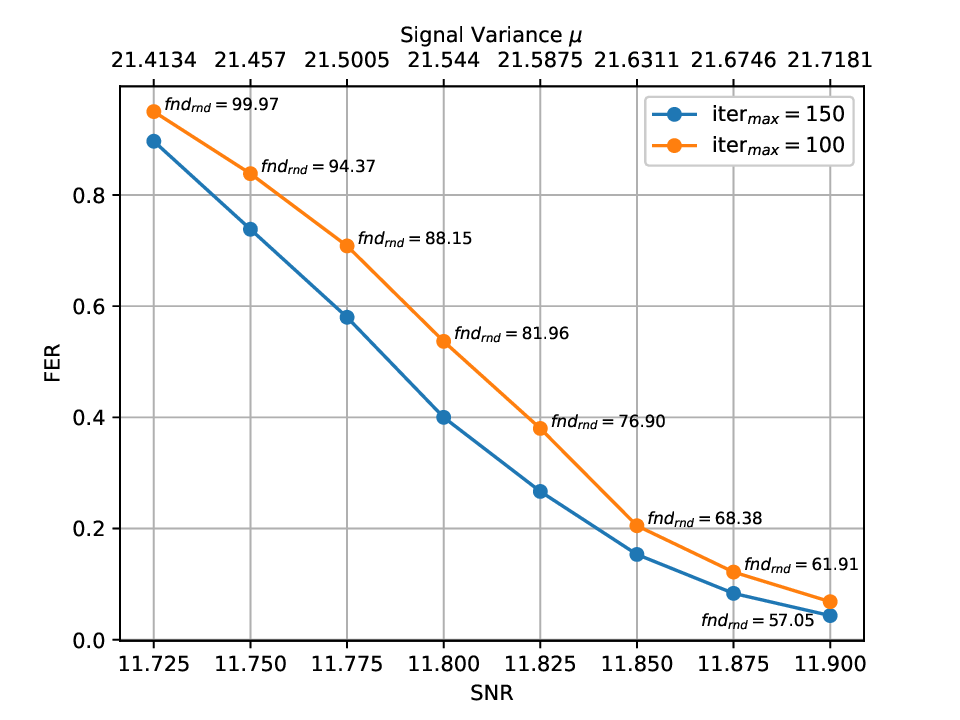}
     \vspace{-0.2cm}
    \caption{FER versus SNR for $p=7$. The FER is compared for the same simulations, when the maximum number of EC iterations is $\text{iter}_\text{max} = 150$ (blue line) and when $\text{iter}_\text{max} = 100$ (orange line). Every point represents the average value of $\text{FER}$, which is obtained after 6 simulations. The step of the SNR is $0.025$. The values of the reconciliation efficiency $\beta$ are chosen so as to produce $R_\text{code} \approx 0.875$. The signal variance $\mu$ that was used to achieve the respective SNR is displayed on the top axis with an accuracy of 4 decimal digits. The average number of iterations $\text{fnd}_\text{rnd}$ needed to decode and verify a block is displayed for every point next to their respective points. The other parameters are constant and listed in Table~\ref{res_param1}. We observe that a slight increase of $\mu$ causes the FER to decline rapidly.}
    \label{fig:FER_vs_SNR}
\end{figure}

\begin{figure}[t]
    \vspace{-0.1cm}
    \centering
    \includegraphics[width=0.45\textwidth]{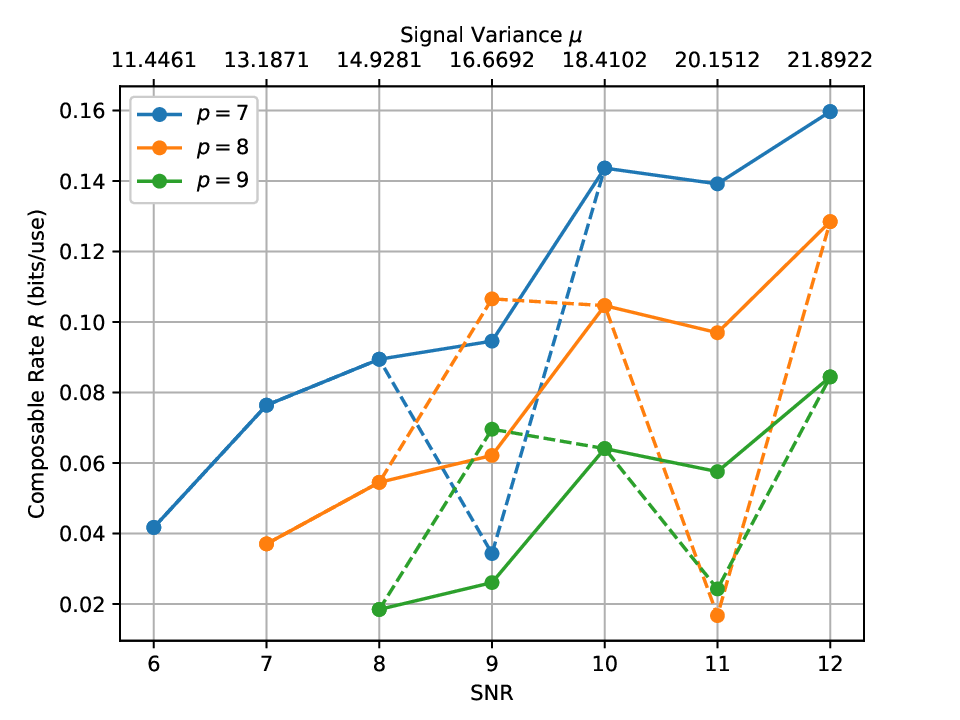}
     \vspace{-0.2cm}
    \caption{Composable secret key rate $R$ versus SNR for discretization bits $p=7,8,9$. The chosen reconciliation efficiency $\beta$, for each value of the $\text{SNR}$, is shown in Table~\ref{recon_efficiencies_choice}. For SNRs = 9 and 11, the solid lines follow the values of the entries `a' of Table~\ref{recon_efficiencies_choice}, while the dashed lines describe  the `b' cases. We observe that, for lower values of $p$ (at a fixed $q=4$), we obtain higher rates for the corresponding $\text{SNR}$. The signal variance $\mu$ that was used to achieve the respective SNR is displayed on the top axis with an accuracy of 4 decimal digits. Other parameters are chosen as in Table~\ref{res_param1}.}
    \label{fig:R_vs_SNR}
\end{figure}

\begin{figure}[t]
    \vspace{-0.1cm}
    \centering
    \includegraphics[width=0.45\textwidth]{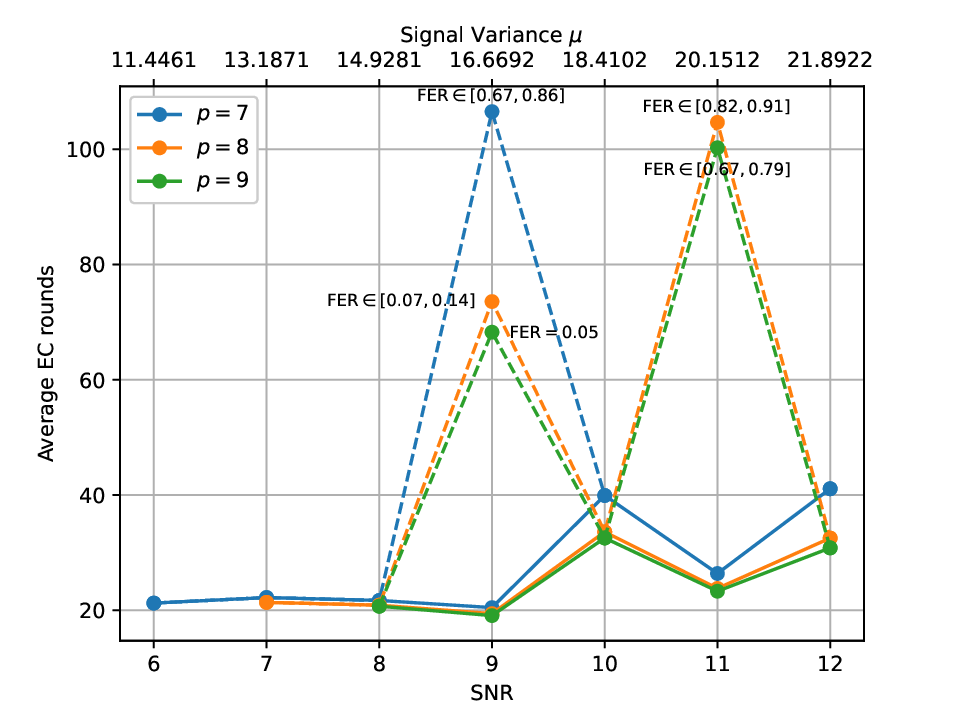}
     \vspace{-0.2cm}
    \caption{Average EC rounds $\text{fnd}_\text{rnd}$ needed to decode a frame versus the SNR for $p=7$, $p=8$ and $p=9$. A round is registered only if the frame passes the verification step. The chosen reconciliation efficiency $\beta$ for each value of the $\text{SNR}$ is shown in Table~\ref{recon_efficiencies_choice}. For $\text{SNR} = 9$ and $\text{SNR} = 11$ specifically, the solid lines respectively follow the values of the entry $9_\text{a}$ and $11_\text{a}$  of Table~\ref{recon_efficiencies_choice}, while the dashed lines describe the $9_\text{b}$ and $11_\text{b}$ cases. For the `b' cases, the FER is reported next to the respective values. The signal variance $\mu$ that was used to achieve the respective SNR is displayed on the top axis with an accuracy of 4 decimal digits. Other parameters are chosen as in Table~\ref{res_param1}.}
    \label{fig:avg_rnd_vs_SNR}
\end{figure}

Figs.~\ref{fig:FER_vs_SNR}, \ref{fig:R_vs_SNR} and~\ref{fig:avg_rnd_vs_SNR} explore different quantities of interest (FER, rate, and EC rounds respectively) as a function of the $\text{SNR}$ and for various choices of the number $p$ of discretization bits. The parameters used in the simulations are given in Table~\ref{res_param1}, where $\mu$ is variable and adapted to attain the desired $\text{SNR}$. In particular, in Figs.~\ref{fig:R_vs_SNR} and~\ref{fig:avg_rnd_vs_SNR}, the reconciliation efficiency $\beta$ (shown in Table \ref{recon_efficiencies_choice}) is chosen according to the following rationale: (i) because a regular LDPC code only achieves a specific value of $R_\text{code}$, $\beta$ is chosen so that $R_\text{code}$ from Eq.~(\ref{RcodeFORMULA0}) matches $R_\text{code}$ of a regular LDPC code with high numerical accuracy; (ii) $\beta$ is high enough so that a positive key rate can be achieved for various values of $p$ for the same $\text{SNR}$; and (iii) $\beta$ is low enough so that a limited number of EC rounds exceeds the iteration limit $\text{iter}_\text{max}$ (if $\beta$ is too high, this limit is exceeded and $\text{FER}$ increases or can even be equal to $1$, meaning that no block is correctly decoded).

Fig.~\ref{fig:FER_vs_SNR} shows the $\text{FER}$ for different values of the $\text{SNR}$. As seen, the $\text{FER}$ is higher for lower $\text{SNRs}$ and quickly declines even with a small increase of the $\text{SNR}$. Note that every simulation, which was executed to produce the particular data, returned a positive key rate (the highest $\text{FER}$ attained was $\text{FER}=0.95$ for $\text{SNR}=11.725$). This result suggests that when $N$ is adequately large, a positive $R$ can be achieved even with a minimal number of correctly decoded and verified blocks. The plot also shows the $\text{FER}$ for the same simulations, if the maximum iteration limit had instead been $\text{iter}_\text{max}=100$. In the case of $\text{SNR}=11.725$, if we had set $\text{iter}_\text{max}=100$, a positive $R$ would not have been realised for some simulations.

Fig.~\ref{fig:R_vs_SNR} shows the composable key rate $R$ versus SNR for different discretization values $p$, while keeping the value of $q$ constant and equal to $4$ (see the list of parameters in Table~\ref{res_param1}). As observed, for fixed values of SNR and $\beta$, the lower the $p$ is, the higher the rate $R$ is. For every $\text{SNR}$ and $\beta$, there is a maximum value for $p$ able to achieve a positive $R$. For example, for $\text{SNR}=6$ and $\beta\approx0.8588$  ($R_\text{code}\approx0.75$) a positive $R$ is impossible to achieve with $p \geq 8$. For $\text{SNR}=7$ and $\beta\approx0.8775$ ($R_\text{code}\approx0.777$), a positive $R$ is infeasible with $p \geq 9$. The key rate improvement owed to smaller values of $p$ relies on the fact that a smaller amount of bits $d=p-q$ are declared publicly, while the protocol maintains a good EC performance thanks to a sufficiently large number of EC iterations. On the other hand, by increasing $p$ for a fixed $q$, we increase the number of the public $d$-bits assisting the LDPC decoding via the sum-product algorithm. This means that the EC step is successfully terminated in fewer rounds.

\begin{table}[t]
\begin{tabular}
[c]{|l|l|l|l||l|l|}
\hline
$\text{SNR}$ & $\beta_{p=7}$ & $\beta_{p=8}$ & $\beta_{p=9}$  & $R_\text{code}$ & $d_{c}$ \\\hline
$6$ & $0.8588$  &  & &    $0.75$                & $8$ \\
$7$ & $0.8788$  & $0.8775$ & & $0.777$          & $9$ \\
$8$ & $0.8868$  & $0.8865$ & $0.8864$ & $0.8$   & $10$ \\
$9_\text{a}$    & $0.89$   & $0.8897$ & $0.8896$ & $0.818$ & $11$ \\
$9_\text{b}$            & $0.9265$   & $0.9262$ & $0.9261$ & $0.833$           & $12$ \\
$10$           & $0.9194$  & $0.9190$ & $0.9189$ & $0.846$            & $13$ \\
$11_\text{a}$  & $0.9116$  & $0.9113$ & $0.9113$ & $0.857$ & $14$ \\
$11_\text{b}$  &  & $0.9327$ & $0.9326$ & $0.866$ & $15$ \\
$12$           & $0.9218$  & $0.9215$ & $0.9214$ & $0.875$ & $16$ \\
\hline
\end{tabular}
 \vspace{0.1cm}
\caption{The chosen reconciliation efficiency $\beta$ for each $\text{SNR}$ of Figs.~\ref{fig:R_vs_SNR} and \ref{fig:avg_rnd_vs_SNR}, together with its respective code rate $R_\text{code}$ and the row weight $d_{c}$ of the LDPC code. A missing value for the reconciliation efficiency implies that the returned composable key rate will most likely be negative under the specified values. The column weight $d_{v}$ remains constant and equal to $2$ for all simulations.}\label{recon_efficiencies_choice}
\end{table}

In Fig.~\ref{fig:avg_rnd_vs_SNR}, we plot the average number of EC rounds $\text{fnd}_\text{rnd}$ required to decode a block versus the SNR, for different values of $p$. 
For a larger value of $p$, fewer decoding rounds are needed. This does not only make the decoding faster, but, depending on the specified $\text{iter}_\text{max}$, it also gives the algorithm the ability to achieve a lower FER. 
Thus, a higher $p$ can potentially achieve a better $p_\text{EC}$, while a smaller $p$ may return a better $R$ (assuming that $\text{iter}_\text{max}$ is large enough). 
Therefore, at any fixed SNR and $\text{iter}_\text{max}$, one could suitably optimize the protocol over the number of discretization bits $p$. 

\section{Conclusions\label{sec:conclusions}}

In this manuscript we have provided a complete procedure for the post-processing of data generated by a numerical simulation
(or an equivalent experimental implementation) of a Gaussian-modulated coherent-state CV-QKD protocol in the high SNR regime. The procedure goes into the details of the various steps of parameter estimation,
error correction and privacy amplification, suitably adapted to match the setting of composable finite-size security. Together
with the development of the theoretical tools and the corresponding technical details, we provide a corresponding Python library
that can be used for CV-QKD simulation/optimization and for the realistic post-processing of data from Gaussian-modulated CV-QKD protocols.

\section*{Acknowledgements}
A. M. was supported by the Engineering and Physical Science Research Council (EPSRC) via a Doctoral Training Partnership EP/R513386/1.

\appendix
\section{Alternative formulas for PE}\label{PEapp}
One may define the estimator for the square-root transmissivity $\tau=\sqrt{\eta T}$ as follows 
\begin{align}\label{eq:Levest}
\widehat{\tau}  &  =\frac{\sum_{i=1}^{m}x_{i}y_{i}}{\sum_{i=1}^{m}x_{i}^{2}}\simeq\frac{1}{m\sigma_x^2}\sum_{i=1}^{m}x_{i}y_{i}.
\end{align}
We then calculate its variance
\begin{align}
\text{Var}(\widehat{\tau})=&\frac{\text{Var}\left(\sum_{i=1}^{m}x_{i}y_{i}\right)}{m^2(\sigma_x^2)^2}=\frac{\text{Var}(xy)}{m(\sigma_x^2)^2}\notag\\
=&\frac{2}{m}\tau^2+\frac{\sigma_z^2}{m\sigma_x^2}
:=\sigma^2_\tau .
\end{align}
Thus the worst-case estimator for the transmissivity $T=\tau^2$ will be given by
\begin{align}
T_m=&\frac{(\tau-w \sigma_\tau)^2}{\eta} =\frac{\left(\sqrt{\eta T}-w\sqrt{\frac{2}{m}\eta T+\frac{\sigma_z^2}{m \sigma_x^2}}\right)^2}{\eta}\notag\\
&=\frac{\eta T-2w\sqrt{\eta T}\sqrt{\frac{2}{m}\eta T+\frac{\sigma_z^2}{m \sigma_x^2}}}{\eta}+\mathcal{O}(1/m)\notag\\
&\simeq T\left(1-2w\sqrt{1/m}\sqrt{2+\frac{\sigma_z^2}{\eta T \sigma_x^2}}\right).\label{USE}
\end{align}
The expression above is the same as the one derived in the main text via Eq.~(\ref{eq:wcvalues})

One may derive a less stringent estimator
by assuming the approximation  $\sum_{i=1}^m x^2_i\simeq  m\sigma_x^2$,
meaning that a sample of size $m$ from the data is close enough to reproduce the theoretical variance $\sigma_x^2$. In such a case, one may write
\begin{align}
\widehat{\tau}&\simeq\frac{1}{m\sigma_x^2}\sum_{i=1}^m x_i(\tau x_i+z_i)=\frac{1}{m\sigma_x^2}\left( \tau \sum_{i=1}^m x^2_i+\sum_{i=1}^m x_i z_i\right)\notag\\
&\simeq \frac{1}{m\sigma_x^2}\left(\tau m\sigma_x^2+\sum_{i=1}^m x_i z_i\right)=\tau+\frac{\sum_{i=1}^m x_i z_i}{m\sigma_x^2}
\end{align}
Therefore the variance is now given by
\begin{align}
\text{Var}(\widehat{\tau})&=\frac{\text{Var}\left(\sum_{i=1}^m x_i z_i\right)}{m^2(\sigma_x^2)^2} \\
&=\frac{\text{Var}\left(xz\right)}{m(\sigma_x^2)^2}=\frac{\sigma_z^2}{m\sigma_x^2}:=(\sigma^\prime_\tau)^2,
\end{align}
yielding the worst-case parameter
\begin{align}\label{eq:Lev}
T^\prime_m=&\frac{(\tau-w\sigma^\prime_\tau)^2}{\eta}\notag\\
\simeq &T\left(1-2w\sqrt{1/m}\sqrt{\sigma^2_z/(\eta T\sigma_x^2)}\right).
\end{align}

We then observe that the relation in Eq.~(\ref{USE}) gives a more pessimistic value for the worst-case transmissivity due to an extra term equal to $2$ appearing in the square root term $\sqrt{2+ \sigma_z^2/(\eta T \sigma_x^2)})$ which is missing in Eq.~(\ref{eq:Lev}). In our main text we assume the most conservative choice corresponding to the estimator in Eq.~(\ref{USE}).

\section{Calculations in $\mathcal{GF}(q)$\label{ap:GLFD}}
A Galois field is a field with finite number of elements. A common way to derive it is to take the modulo of the division of the integers over a prime number $p$. The order of such a field $q=p^k$ (with $k$ being a positive integer) is the number of its elements. 
All the Galois fields with the same number of elements are isomorphic and can be identified by $\mathcal{GF}(q)$. A special case is the order $q=2^k$. In a field with such an order, each element is associated with a binary polynomial of degree no more than $k-1$, i.e. the elements can be described as $k$-bit strings where each bit of the string corresponds to the coefficient of the polynomial at the same position. For instance, for the element $5$ of $\mathcal{GF}(2^3)$ we have
$$101\rightarrow x^2+1.$$ This is instructive on how the operations of addition and multiplication are computed in such a field. For example, the addition of $5$ and $6$ is made in the following way
$$101+110\rightarrow(x^2+1)+(x^2+x)=\underbrace{\overbrace{(1+1)}^0 x^2+x+1}_{011 \rightarrow 3}.$$ As the field is finite, one can also perform the addition using a precomputed matrix. For instance, for $\mathcal{GF}(2^3)$, we have
\begin{equation}\label{eq:Apc}
\mathbf{A}_3 = \left(
\begin{array}
[c]{cccccccc}%
   0  &  1 &  2 &  3 &  4 &  5 &  6 &  7 \\
   1  &  0 &  3 &  2 &  5 &  4 &  7 &  6 \\
   2  &  3 &  0 &  1 &  6 &  7 &  4 &  5 \\
   3  &  2 &  1 &  0 &  7 &  6 &  5 &  4 \\
   4  &  5 &  6 &  7 &  0 &  1 &  2 &  3 \\
   5  &  4 &  7 &  6 &  1 &  0 &  3 &  2 \\
   6  &  7 &  4 &  5 &  2 &  3 &  0 &  1 \\
   7  &  6 &  5 &  4 &  3 &  2 &  1 &  0
\end{array}
\right).
\end{equation}

Subtraction between two elements of $\mathcal{GF}(2^k)$ gives the same result as addition, making the two operations equivalent. Multiplication is more complicated, especially when the result is a polynomial with a degree larger then $k-1$. For example, in $\mathcal{GF}(2^3)$, $7 \times 6$ is calculated as
\begin{align}
111 \times 110 &\rightarrow (x^2+x+1)\times(x^2+x)\notag\\
=&x^4+x^3+x^3+x^2+x^2+x=x^4+x.
\end{align}
Because we have a degree $4$ polynomial, we need to take this result modulo an irreducible polynomial of degree $3$, e.g., $x^3-x+1$. 
Thus, we have
\begin{equation}\label{eq:modulo}
(x^4+x~\text{mod}~x^3-x+1 )=x^2\rightarrow 100\rightarrow 4,
\end{equation}
where the operation can be made by adopting a long division with exclusive OR~\cite{GaloisBook}.
Instead, as seen in addition, multiplication can be performed by using a precomputed matrix. For instance, in $\mathcal{GF}(2^3)$, the results are specified by the following matrix
\begin{equation}\label{eq:Ppc}
\mathbf{M}_3 = \left(
\begin{array}
[c]{cccccccc}%
   0 &  0 &  0 &  0 &  0 &  0 &  0 &  0 \\
   0 &  1 &  2 &  3 &  4 &  5 &  6 &  7 \\
   0 &  2 &  4 &  6 &  3 &  1 &  7 &  5 \\
   0 &  3 &  6 &  5 &  7 &  4 &  1 &  2 \\
   0 &  4 &  3 &  7 &  6 &  2 &  5 &  1 \\
   0 &  5 &  1 &  4 &  2 &  7 &  3 &  6 \\
   0 &  6 &  7 &  1 &  5 &  3 &  2 &  4 \\
   0 &  7 &  5 &  2 &  1 &  6 &  4 &  3
\end{array}
\right).
\end{equation}
\section{LDPC decoding\label{ap:LDPC decoding}}
\subsection{Updating the likelihood function}
Let us assume a device where its output is described by the random variable $X$ taking values $x$ according to a family of probability distributions $\mathcal{P}(X;\theta)$ parametrized by $\theta$. Given the sampled data string $X_i$ for $i=1,\dots,n$ from this distribution, one can build a string of data $X^n$ and define the likelihood of a specific parameter $\theta$  describing the associated probability distribution as
\begin{equation}
\mathcal{L}(\theta|X^n)=P(X^n|\theta)=\prod_{i=1}^n P(X_i|\theta),
\end{equation}
where $P(X^n|\theta)$ is the conditional probability for a specific $X^n$ to come out of the device given that its distribution is described by $\theta$ and the its outcome of the device is i.i.d. following $\mathcal{P}(X;\theta)$. Intuitively, a good guess $\widehat{\theta}$ of the parameter $\theta$ would be the argument $\theta^*$ of the maximization of the likelihood function over $\theta$.
Using Bayes' rule, we may write
\begin{equation}
P(X^n|\theta)=\frac{P(X^n)}{P(\theta)}P(\theta|X^n),
\end{equation}
and observe that $P(X^n)$ is not dependent on $\theta$ and $P
(\theta)$ is considered uniform (thus independent of $\theta$). Therefore, one may maximize $P(\theta|X^n)$ instead. Furthermore, for the simplification of the later discussion one may express the previous probability as a function being only dependent from the parameter $\theta$ (considering the data $X_i$ as constants), namely
\begin{equation}
    F(\theta)=P(\theta|X^n).
\end{equation}

Let us now consider the case where we have a vector of parameters (variables) $\vec{\theta}=(\theta_1,\dots,\theta_n)$ describing the distribution $\mathcal{P}(X|\boldsymbol{\theta})$. Respectively, one can define the probability
\begin{equation}\label{eq:f_function}
F(\vec{\theta})=F(\theta_1,\dots,\theta_n),
\end{equation}
and its marginals
\begin{equation}\label{eq:f_marginals}
F(\theta_i)=\sum_{k\neq i}F(\theta_1,\dots,\theta_k,\dots,\theta_n).
\end{equation}

\begin{table}[b]
    \centering
    \begin{tabular}{c}
  $\mathbf{H}= \left[\begin{array}{c c c c c}
      0 &0&3&0&1  \\
      2 &0&0&1&0 \\
      0 &1&0&2&3
        \end{array}\right]$
    \end{tabular}
    \caption{An example for a $l\times n$ parity check matrix with values in $\mathcal{GF}(2^2)$ for $l=3$ checks (check nodes) and $n=5$ transmitted signals (variable nodes). For this matrix, the assumptions of a regular code explained in Sec.~\ref{sec:EC} are not valid and it is used only as a toy model for the convenience of the description for the sum-product algorithm.}
    \label{tab:pcmexample}
\end{table}

Let us now assume that there are certain constraints that $\vec{\theta}$ should satisfy which are summarized by a system of $m$ linear equations (checks) $\mathbf{H}\vec{\theta}=\vec{z}$, where $\mathbf{H}$ is an $m\times n$ matrix. In particular, there are $m$ equations that the $\theta_i$ should satisfy in the form of
\begin{equation}
\sum_i\mathbf{H}_{ji}\theta_i=z_j~~\text{for}~~ j=1,\dots,m.
\end{equation}
For instance, when $\vec{z}=(3,1,2)$, the matrix in Table~\ref{tab:pcmexample} gives the following three equations~\cite{Note_GF}:
\begin{align}
3 \theta_3+ \theta_5=&3,\\
2\theta_1+\theta_4=&1,\\
\theta_2+2 \theta_4+3\theta_5=&2.
\end{align}
Then one needs to pass from the (a priori) probability distribution of Eq.~(\ref{eq:f_function}) to
\begin{equation}
\widetilde{F}(\vec{\theta})=F(\vec{\theta}|\mathbf{H}\vec{\theta}=\vec{z}),
\end{equation}
and calculate the respective marginals
\begin{equation}\label{eq:ftilde_marginals}
\widetilde{F}(\theta_i)=F(\theta_i|\mathbf{H}\vec{\theta}=\vec{z}).
\end{equation}

\begin{figure}[t]
\vspace{0.1cm}
\includegraphics[width=0.4\textwidth]{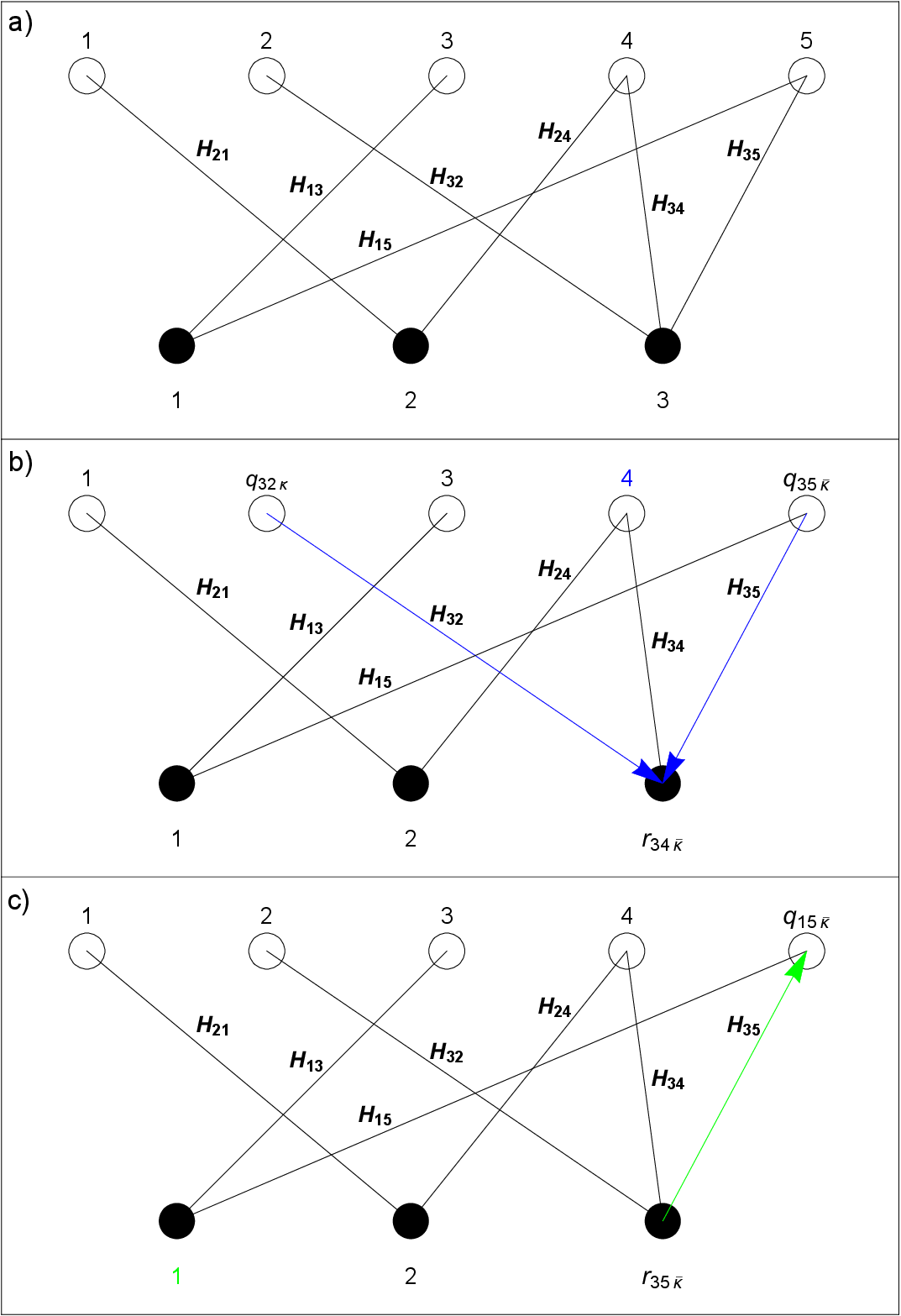}
\caption{a) Tanner graph of the parity check matrix of Table~\ref{tab:pcmexample}. The variable (output) nodes (white disks) are connected with the check (syndrome) nodes (black disks) when $\mathbf{H}_{ji} \neq 0$. b) One instance of the horizontal step (Step 2) of Algorithm~\ref{algo:IBPA}. Here, the signal probability $r_{34\overline{\kappa}}$ is updated for all the  $\overline{\kappa} \in \mathcal{GF}(2^2)$ from the contribution (blue arrows) of the rest of the neighbour variable nodes of check node $3$, apart from the variable node $4$ (node in blue). This update will be repeated in the same step for all the variable nodes, i.e., $r_{32\overline{\kappa}}$ and $r_{35\overline{\kappa}}$ will be calculated too. The same procedure will be followed for syndrome nodes $1$ and $2$ before the algorithm passes to the horizontal step. This description provides the conceptual steps to derive the desirable result. Practically, the algorithm follows a more complex path, e.g., calculates probabilities of partial sums. However, this path gives an advantage in terms of efficient calculations.  c) An instance of the horizontal step (Step 3) of Algorithm~\ref{algo:IBPA}. Here the probability $q_{15\overline{\kappa}}$ is updated for all the $\overline{\kappa}\in \mathcal{GF}(2^2)$. It is updated only from the contribution of syndrome node $3$ (green arrow), while node $1$ (node in green) is not participating. This update will happen for all the syndrome nodes, namely $q_{35\overline{\kappa}}$ will be calculated too. It will be repeated also for all variable nodes, before the tentative decoding (Step 4) is going to start.}
\label{fig:tanner}
\end{figure}

\begin{algorithm}
\caption{\label{algo:IBPA}Non-Binary Sum-Product Algorithm}
\textbf{Input:}$P(\overline{K}_i|X_i \underline{K}_i),K^{l}_\text{\text{sd}}$, \textbf{Output:}$\widehat{K}^{n}, \text{fnd}, \text{fnd}_\text{rnd}$
\begin{algorithmic}[1]
\State \textbf{Step 1:} Initialization
		\State $\vec{z} \leftarrow K^{l}_\text{\text{sd}}$
		\State $j,i \leftarrow j,i: \mathbf{H}_{ji}\neq0$ (Tanner graph creation)
  	    \State $f_{i}^{\overline{\kappa}} \leftarrow P(\overline{K}_i=\overline{\kappa}|X_i,\underline{K}_i)$
		\State $q_{ji\overline{\kappa}} \leftarrow f_{i}^{\overline{\kappa}}$
		\For {$\text{iter}=1,2,\ldots \text{iter}_\text{max}$}
		\State \textbf{Step 2:} Horizontal Step
        \State $r_{ji\overline{\kappa}} \leftarrow \underset{s, t :~ s+ t = z_{j} - \mathbf{H}_{ji}\overline{\kappa}~~~~~~~~~~~~~~~~~~~~~~~~~}{ \sum\Pr \Big[\sigma_{j(i-1)} = s \Big]\Pr\Big[\rho_{j(i+1)}=t\Big]}$
        \State \textbf{Step 3:} Vertical Step
        \State $q_{ji}^{\overline{\kappa}} \leftarrow \alpha_{ji}f_{i}^{\overline{\kappa}} \prod\limits_{m  \setminus j}r_{mi\overline{\kappa}}$, $\alpha_{ji}: \sum\limits_{\overline{\kappa}=0}^{2^q-1} q_{ji\overline{\kappa}} = 1$
        \State \textbf{Step 4:} Tentative Decoding
        \State $\widehat{K}_{i} \leftarrow \underset{\overline{\kappa}}\argmax{f_{i\overline{\kappa}} \prod\limits_{j}r_{ji\overline{\kappa}}}$
        \If{$\mathbf{H} \widehat{K}^{n}=\vec{z}$}
        \State \Return $\widehat{K}^{n},\text{fnd}_\text{rnd}, \text{fnd} \leftarrow$ True \EndIf
        \If{$\text{iter}=\text{iter}_\text{max}$}
        \State \Return $\text{fnd} \leftarrow$ False \EndIf
		\EndFor
\end{algorithmic}
\end{algorithm}

\subsection{Sum-product algorithm}
The sum-product algorithm uses the intuition of the previous analysis to efficiently calculate the marginals
\begin{equation}\label{eq:ftilde_marginals_K}
\widetilde{F}(\overline{K}_i)=F(\overline{K}_i|\mathbf{H}\overline{K}^n=K_\text{sd}^l)
\end{equation}
of Eq.~(\ref{eq:ftilde_marginals}) for $\theta_i:=\overline{K}_i$, $\vec{\theta}:=\overline{K}^n$, $\vec{z}:=K_\text{sd}^l$ and the a priori marginal probabilities $F(\overline{K}_i=\overline{\kappa})=P(\overline{\kappa}|X_i \underline{K}_i)$, calculated in Eq.~(\ref{eq:a-priori prob}). To do so, it associates a Tanner (factor) graph to the matrix $\mathbf{H}$ and assumes signal exchange between its nodes. More specifically, the graph consists of two kinds of nodes: $n$ variable nodes representing the parameters (variables) $\overline{K}_i$ and $m$ check nodes representing the linear equations (checks) described by $\mathbf{H}\overline{K}^n=K_\text{sd}^l$. Then, for each variable $i$ that participates in the $j$th equation, there is an edge connecting the relevant nodes. At this point, we present an example of such a Tanner (factor) graph in Fig.~\ref{fig:tanner}(a), based on the matrix $\mathbf{H}$ in Table~\ref{tab:pcmexample}. The signal sent from the variable node $i$ to a factor node $j$ is called $q_{ji\overline{\kappa}}$ and is the probability that the variable $\overline{K}_i=\overline{\kappa}$ and all the linear equations are true, apart from equation $j$. The signal sent from the check node $j$ to the variable node $i$ is called $r_{ji\overline{\kappa}}$ and is equal with the probability of equation $j$ to be satisfied, if the variable $\overline{K}_i=\overline{\kappa}$. Note that, based on these definitions, the marginals of Eq.~(\ref{eq:ftilde_marginals_K}) are given by
\begin{equation}\label{eq:tentative}
\widetilde{F}(\overline{K}_i=\overline{\kappa})=q_{ji\overline{\kappa}}r_{ji\overline{\kappa}},
\end{equation}
for any equation $j$ that the variable $i$ takes part in.

In particular, in each iteration, the algorithm updates the $r_{ji\overline{\kappa}}$ (horizontal step) through the signals of the neighbour variable nodes apart from the signal from node $i$  by the following rule: given a vector $\overline{K}^n$ where its $i$th element is equal to $\overline{K}_i=\overline{\kappa}$ we have
\begin{equation}
r_{ji\overline{\kappa}}=\sum_{\{i\}}\text{Prob}[{K_\text{sd}}_j|\overline{K}^n]\prod_{k \in \mathcal{N}(j)\setminus i}q_{jk\overline{K}_k},
\end{equation}
where $\text{Prob}[{K_\text{sd}}_j|\overline{K}^n]$ takes the value $1$, if the check $j$ is satisfied from $\overline{K}^n$, or $0$ if it is not. Note that  the values of $q_{jk\overline{K}_k}$ are initially updated with the a priori probabilities during the initialization step (see line~5  of Algorithm~\ref{algo:IBPA}) and that $\mathcal{N}(j)$ are the set of neighbours of
the $j$th check node. An example of such an update is depicted in Fig.~\ref{fig:tanner}(b).

In fact, the algorithm takes advantage of the fact that
\begin{equation}\label{eq:r_with _partialsums}
r_{ji\overline{\kappa}}=\text{Prob}\left[\sigma_{j (i-1)}+\rho_{j(i+1)}={K_\text{sd}}_j -\mathbf{H}_{ji} \overline{K}_i\right],
\end{equation}
where
\begin{align}
\sigma_{jk}=\sum_{i:i\leq k}\mathbf{H}_{ji}\overline{K}_i,~~ \rho_{jk}=\sum_{i:i\geq k}\mathbf{H}_{ji}\overline{K}_i
\end{align}
are partial sums with different direction running over the $j$th check. More specifically, Eq.~(\ref{eq:r_with _partialsums}) can be further simplified into a sum of a product of probabilities of the previous partial sums taking specific values by satisfying the $j$th check, as in line~$10$ of the pseudocode of Algorithm~\ref{algo:IBPA}. Then, the algorithm updates the $q_{ji\overline{\kappa}}$ through the signals coming from the neighbour check nodes apart from node $j$, as depicted for the example in Fig.~\ref{fig:tanner}(c). The rule to do so is given in line~12 of the pseudocode (vertical step). Finally, in the tentative decoding step, the algorithm takes the product of $q_{ji\overline\kappa}$ and $r_{ji\overline\kappa}$, then calculates and maximizes the marginal of Eq.~(\ref{eq:tentative}) over $\overline{\kappa}$. The arguments $\widehat{K}_i$ of this maximization of every marginal  create a good guess $\widehat{K}^n$ for $\overline{K}^n$. In the next iteration, the algorithm follows the same steps, using the previous $q_{ij\overline{\kappa}}$ to make all the updates.


\section{Toeplitz matrix calculation with Fast-Fourier Transform\label{ap:toeplitz_sk}}
The time complexity of the dot product between a Toeplitz matrix $\mathbf{T}$ and a sequence $\mathbf{x}$ is $O(\tilde{n}^{2})$. This complexity can be reduced to $O(\tilde{n}\log{\tilde{n}})$ using the definition of a circulant matrix and the FFT. A circulant matrix $\mathbf{C}$ is a special case of the Toeplitz matrix, where every row of the matrix is a right cyclic shift of the row above it~\cite{CirculantMatrixRev}. Such a matrix is always square and is completely defined by its first row $\mathbf{C}_\text{def}$.

\begin{figure*}[t!]
\centering
\begin{minipage}[b]{.45\textwidth}
\includegraphics[width=1\textwidth]{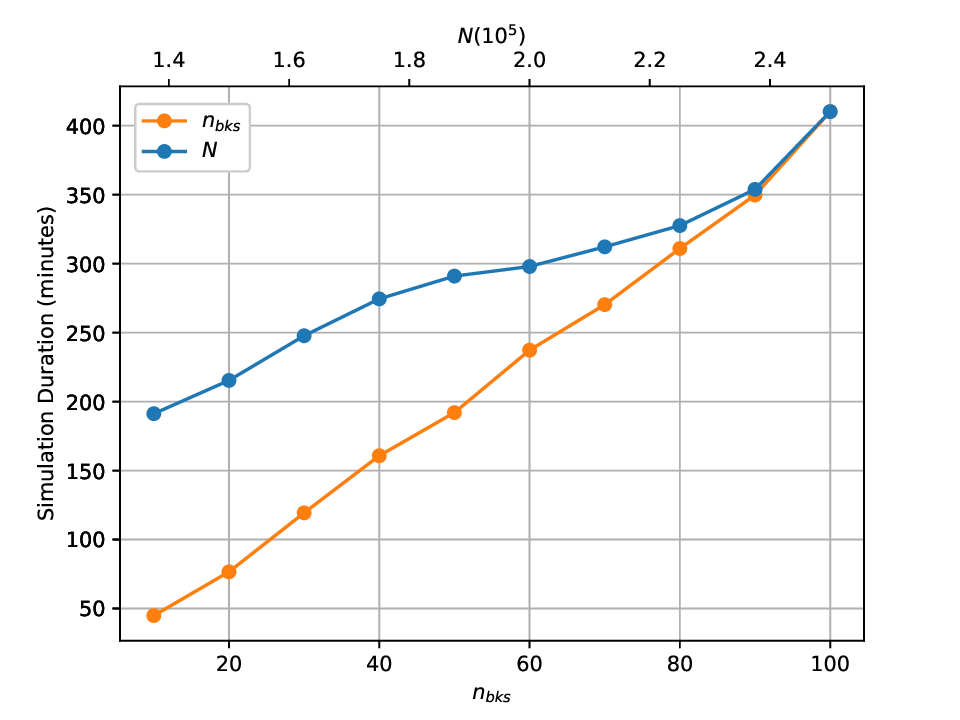}
\end{minipage}\qquad
\begin{minipage}[b]{.45\textwidth}
\includegraphics[width=1\textwidth]{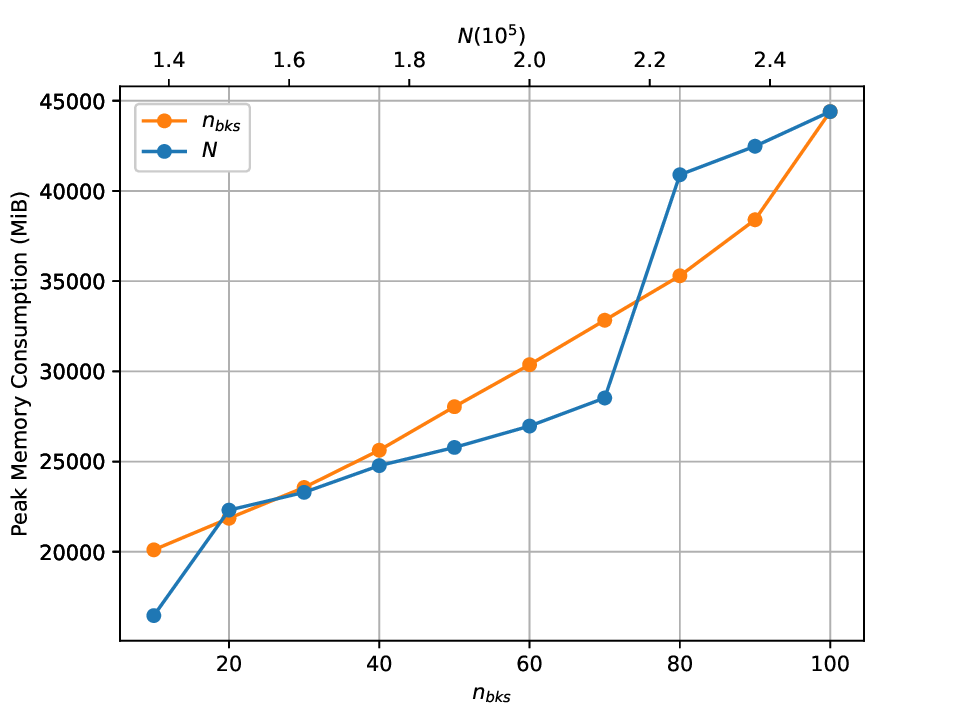}
\end{minipage}
\vspace{-0.3cm}
\caption{\label{benchmarks}Computational benchmarks: Entire duration of a simulation (left) and peak memory consumption (right). These are provided with respect to a variable block size $N$ (in blue) and a variable number of blocks $n_\text{bks}$ (in orange). When $N$ is variable, the number of blocks is constant and equal to $n_\text{bks}=100$. When $n_\text{bks}$ is variable, the block size is constant and equal to $N=250000$. The results depicted are based on the simulations of Fig.~\ref{fig:R_vs_N} and Fig.~\ref{fig:R_vs_n_bks}, respectively (whose corresponding sets of parameters can be found in Table \ref{res_param1}). In order to successfully decode a block, around $40$ iteration rounds are required on average.}
\end{figure*}

The steps are as follows~\cite{privacy_amplification_steps}:
\begin{itemize}
    \item The Toeplitz matrix is reformed into a circulant matrix by merging its first row and column together. Since the former has dimensions $\tilde{n} \times r$, where $\tilde{n}$ is the privacy amplification block length and $r$ is the length of the final key, the length of the definition of the latter becomes $\tilde{n} + r - 1$.
    \item The decoded sequence to be compressed is extended, as $r - 1$ zeros are padded to its end. The length of the new sequence $\mathbf{S}_\text{ext}$ is now equal to the length of the circulant matrix definition.
    \item To efficiently calculate the key, an optimized multiplication is carried out as
    \begin{equation}
        \mathfrak{F}^{-1}[\mathfrak{F}(\mathbf{S}_\text{ext}) * \mathfrak{F}(\mathbf{C}_\text{def})]
    \end{equation}
    where $\mathfrak{F}$ represents the FFT and $\mathfrak{F}^{-1}$ stands for the inverse FFT. Because of the convolution theorem, the $*$ operator signifies the Hadamard product and therefore element-wise multiplication can be performed.
    \item As the key format is required to be in bits, the result of the inverse FFT is taken modulo $2$.
    \item The key $\mathbf{K}$ is constituted by the first $r$ bits of the resulting bit string of length $\tilde{n} + r - 1$.
\end{itemize}

\section{Software benchmarks}

The benchmarks for the entire runtime duration and the peak memory consumption, with regards to different sizes for the $N$ and $n_\text{bks}$ variables, are portrayed in Fig.~\ref{benchmarks}. Around $95\%$ of the runtime is ascribed to the duration of the non-binary sum-product algorithm, while the heavy memory load is predominantly because of the privacy amplification stage. The slow speed of the decoding stage is justified, as the non-binary sum-product algorithm is complex in nature. In addition, in order to achieve a positive composable key rate, the block sizes have to be sufficiently large ($N>10^5$) and an adequate number of blocks has to be present as well. 

For such a computational task, we employed the Interactive Research Linux Service of University of York, whose specifications are noted in Table \ref{specs}. Nevertheless, the software is able to run on a conventional computer as well; however, the speed will be significantly diminished. To provide algorithmic speedups we used various techniques, which include, but are not limited to the Numba library, parallelization, the use of dictionary structures (whose lookup time complexity is $\mathcal{O}(1)$) and precomputed tables for the Galois field computations.

\begin{table}[h]
\begin{tabular}
[c]{|c|c|c|c|}
\hline
CPU Model & Intel Xeon E5-2680 v4   \\
CPU Clock Speed & 2.60 GHz  \\
Number of Cores & 56   \\
RAM & 512GB  \\
OS & Ubuntu 20.04   \\
Python Version & 3.8 \\
\hline
\end{tabular}
\vspace{0.1cm}
\caption{The specifications of the system, on which the simulations were executed.}\label{specs}
\end{table}

An advantage of the sum-product algorithm is that it is parallelizable. Therefore, possessing more processing cores is beneficial in terms of speed and, consequently, such projects are often carried out on Graphical Processing Units (GPUs)~\cite{Mario} because of their superior number of cores compared to Central Processing Units (CPUs). To provide massive compatibility, the software is written to target solely CPUs. Future versions of the software may process the error correction stage on a GPU level. In addition, the non-binary sum-product method used in this paper is anachronistic in terms of speed. There exists a newer method, which replaces the stages that demand the most complexity with FFT computations~\cite{fft_LDPC}. Future improvements on the algorithm could potentially include this method as well.

Generating a shared secret key $\mathbf{K}$ for a particular set of noise parameters in a quick manner is a matter of optimization of the block size, the number of blocks and the discretization bits. Let us examine the well-studied case of $\text{SNR}=12$. In Sec. \ref{sec:simulations}, it is explained that selecting a small number of blocks with a large block size over a large number of blocks with a smaller block size is beneficial for the key rate. Furthermore, in Fig. \ref{fig:avg_rnd_vs_SNR}, it is shown that $p=8$ provides a reasonable boost in the error-correction speed. On the contrary, choosing $p=9$ offers little advantage when the average number of iterations is already small; however the key rate sacrifice is large. Under $R_\text{code}=0.875$ and $p=8$, the algorithm needs around $32.5$ iterations on average to decode a block. Taking all the above into consideration, a simulation with the parameters $N=352000$, $n_\text{bks} = 5$ and $p=8$ can produce a key with a composable key rate of around $10^{-4}$ bits/use and a delay time of less than 30 minutes.


\begin{thebibliography}{99}
\bibitem{library} The Python library is available on github at the address softquanta/homCVQKD
\bibitem{noteI} Note that $\alpha$ and $q$ also affect the a priori probabilities of the decoding step in Eqs.~(\ref{eq:a-priori prob}) and~(\ref{eq:conditional prob}) which, in turn, affect the $p_\text{EC}$ for a given number of maximum iterations $\text{iter}_\text{max}$.
\bibitem{noteRATE}We remark that the formula for the key rate in Eq.~(\ref{rateCOMPO}) is valid under the assumption of collective Gaussian attacks, which is the state-of-the-art highest level of security for the Gaussian-modulated homodyne protocol. Security of the heterodyne protocol~\cite{noswitch} could be extended to fully coherent attacks due to the extra symmetry of the protocol under the unitary group provided by the hetorodyne measurement~\cite{revQKD}. In order to enforce this symmetry, the gathered data associated with block size $N$ needs to be multiplied by a random orthogonal $N\times N$ matrix. To the best of our knowledge, the complexity for creating such a matrix is $\mathcal{O}(N^2\log N)$~\cite{Stewart_random_orthogonal}. While this is certainly possible from a theoretical point of view, it is very challenging to be implemented in practical data processing since the value of the  block size is $N > 10^5$. The time overhead involved in this step is extremely long (with current processing power).
\bibitem{Note_GF}{Note that these equations are valid in $\mathcal{GF}(4)$}
\end{thebibliography}
\end{document}